\documentclass[onecolumn,useamsfonts]{pasj00}
\usepackage{epsf}
\usepackage{graphicx}
\usepackage{amssymb} 
\usepackage{float} 
\usepackage{lscape} 
\usepackage{rotating} 
\usepackage[mathscr]{eucal}
\usepackage{color} 
\newcommand\simlt{\hspace{0.3em}\raisebox{0.4ex}{$<$}\hspace{-0.75em}\raisebox{-.7ex}{$\sim$}\hspace{0.3em}} 
\newcommand\simgt{\hspace{0.3em}\raisebox{0.4ex}{$>$}\hspace{-0.75em}\raisebox{-.7ex}{$\sim$}\hspace{0.3em}} 

\begin{document}
\SetRunningHead{Kandori et al.}{IR Polarimetry of NGC 2024}
\Received{2006/11/7}
\Accepted{2007/1/29}

\title{Near-Infrared Imaging Polarimetry of the Star Forming Region NGC 2024}

\author{Ryo \textsc{KANDORI},\altaffilmark{1} 
        Motohide \textsc{TAMURA},\altaffilmark{1, 2} 
        Nobuhiko \textsc{KUSAKABE}\altaffilmark{2}
		Yasushi \textsc{NAKAJIMA},\altaffilmark{1}} 
\author{Takahiro \textsc{NAGAYAMA},\altaffilmark{5} 
		Chie \textsc{NAGASHIMA},\altaffilmark{3} 
		Jun \textsc{HASHIMOTO},\altaffilmark{1, 3} 
        Akika \textsc{ISHIHARA},\altaffilmark{2}} 
\author{Tetsuya \textsc{NAGATA},\altaffilmark{5} 
\and
		Jim \textsc{HOUGH}\altaffilmark{6}} 
		  
\altaffiltext{1}{National Astronomical Observatory, 2-21-1 Osawa, Mitaka, Tokyo 181-8588; kandori@optik.mtk.nao.ac.jp}
\altaffiltext{2}{Graduate University of Advanced Science, 2-21-1 Osawa, Mitaka, Tokyo 181-8588}
\altaffiltext{3}{Department of Physics, Tokyo University of Science, 1-3, Kagurazaka, Sinjuku-ku, Tokyo 162-8601}
\altaffiltext{4}{Department of Astrophysics, Nagoya University, Chikusa-ku, Nagoya 464-8602}
\altaffiltext{5}{Department of Astronomy, Kyoto University, Sakyo-ku, Kyoto 606-8502}
\altaffiltext{6}{Centre for Astrophysics Research, University of Hertfordshire, Hatfield HERTS AL10 9AB, UK}

\KeyWords{circumstellar matter --- infrared: stars ---
ISM: individual (NGC 2024) --- polarization ---
stars: formation} 

\maketitle

\begin{abstract}
We conducted wide-field $JHK_{\rm s}$ imaging polarimetry toward NGC 2024, which is a massive star-forming region in the Orion B cloud. We found a prominent and extended polarized nebula over NGC 2024, and constrained the location of illuminating source of the nebula through the analysis of polarization vectors. 
A massive star, IRS 2b with a spectral type of O8 $-$ B2, is located in the center of the symmetric vector pattern. 
Five small polarized nebulae associated with YSOs are discovered on our polarization images. These nebulae are responsible for the structures of circumstellar matter (i.e., disk/envelope systems) that produce strongly polarized light through dust scattering. 
For the point-like sources, we performed software aperture polarimetry in order to measure integrated polarizations, with which we detected candidate sources associated with circumstellar material by selecting sources with a larger polarization than that estimated from the extinction of foreground material. 
We investigated the fraction of highly polarized sources against the intrinsic luminosity of stars ($\propto$ mass), and found that the source detection rate remains constant from low (brown dwarfs) to higher luminosity (solar-type) stars. This result indicates the relative disk scale-hight is rather independent of the stellar mass. 
We confirmed the result using our polarimetry of the stars with known spectral types in NGC 2024. We found five young brown dwarfs with highly polarized integrated emission. These sources serve as direct evidence for the existence of disk/envelope system around brown dwarfs. 
We investigated the magnetic field structure of NGC 2024 through the measurements of dichroic polarization. The average position angle of projected magnetic fields across the region is found to be 110${}^{\circ}$. 
We found a good consistency in magnetic field structures obtained using near-infrared dichroic polarization and sub-mm/far-infrared dust emission polarization, indicating that the dichroic polarizations at near-infrared wavelengths trace magnetic field structures inside dense ($A_{V} \simlt 50$ mag) molecular clouds. 
\end{abstract}

\section{INTRODUCTION}
NGC 2024 is a massive star forming region in the Orion B giant molecular cloud at a distance of 415 pc (Anthony-Twarog 1982). In the optical images, NGC 2024 shows a bright nebulosity with a prominent central dark dust lane located in front of the ionized gas. Imaging observations at the near-infrared wavelengths clearly revealed an embedded cluster in this region (e.g., Barnes et al. 1989; Lada et al. 1991; Haisch et al. 2001), which is high in stellar density ($\sim 400$ stars pc${}^{-3}$, Lada 1999) and is young in mean age (0.3 Myr, Meyer 1996). 
The structure of the HII region toward NGC 2024, forming at the edge of a molecular cloud, was studied in detail (Crutcher et al. 1986; Barnes et al. 1989). 
Though the dominant source of ionization powering the HII region has not been identified until recently due to heavy obscuration by dust around the central region, Bik et al. (2003) suggested that the dominant source is IRS 2b located 5$''$ north-west of IRS 2. They found that the spectral type of IRS 2b is in the range O8 V $-$ B2 V, which is consistent with the intensities of radio continuum emission and recombination lines observed toward the HII region (Krugel et al. 1982; Barnes et al. 1989). NGC 2024 includes a number of protostars forming along the $\lq \lq$star forming ridge$"$, a filamentary shaped dense molecular material, extending in the north-south direction. In the ridge, dense dust condensations, FIR 1-7, were reported based on the 1.3 mm and 870 $\mu$m observations (Mezger et al. 1988, 1992). 
\par
The magnetic field structure of NGC 2024 was studied by observing linearly polarized thermal emission from aligned dust grains ($B_{\perp}$ measurements) and the Zeeman splitting of molecular or atomic lines ($B_{\parallel}$ measurements). Hildebrand et al. (1995) and Dotson et al. (2000) reported far-infrared (100 $\mu$m) polarization map covering $\sim 3' \times 3'$ toward NGC 2024. They found that the overall magnetic field orientation is in the east-west direction. There is a local magnetic field structure at and around the star forming ridge, but its detailed structure could not be resolved on their polarization vector map. Matthews, Fiege, \& Moriarty-Schieven (2002) studied the magnetic field structure of the star forming ridge in NGC 2024 based on the sub-mm dust continuum observations using the polarimetry mode of JCMT/SCUBA. They revealed the detailed magnetic field structure along the ridge, which is consistent with the previous 100 $\mu m$ lower resolution polarization map. On the basis of their polarization map at 850 $\mu m$, they modeled the magnetic field structure of the star forming ridge as helical magnetic fields surrounding a curved filamentary cloud, and as magnetic field swept by the ionization front of the expanding HII region. 
\par
Crutcher et al. (1999) performed circular polarimetry toward NGC 2024 with the Very Large Array (VLA) to measure the Zeeman splitting of the absorption lines of HI and OH. They found $B_{\parallel}$ smoothly varies from 0 $\mu$G in the northeast to $\sim 100 \mu$G in the southwest of the star forming ridge, and suggested that the orientation of magnetic fields toward the lines of sight varies across the region. 
\par
At the near-infrared wavelengths, polarimetry of a few bright stars in NGC 2024 were reported (IRS 1: Wilking et al. 1980, IRS 1,2: Heckert \& Zeilik 1981). More recently, Moore \& Yamashita (1994) conducted imaging polarimetry toward the infrared nebula associated with FIR 4. However, the field of view of their map was too small ($\sim 40'' \times 40''$) to reveal polarization distribution over the entire surface of NGC 2024. 
\par
In this paper, we present the polarization images of the NGC 2024 star forming region, which is a part of our ongoing project of $JHK_{\rm s}$ polarimetry of star forming regions. Our observations are sufficiently deep ($J=19.2$ mag, ${\rm S}/ {\rm N} = 5$) and wide ($7.7' \times 7.7'$) covering a large extent of  NGC 2024. Near-infrared imaging polarimetry is particularly useful to trace polarized light from star forming regions compared with that at optical wavelengths where dust extinction is severe. 
Our wide-field $JHK_{\rm s}$ polarization images can be used to reveal large-scale infrared reflection nebulae (IRNe); to detect smaller-scale IRNe associated with young stars which is responsible for the circumstellar material (e.g., disk/envelope system), producing strongly polarized light through dust scattering; and to delineate the magnetic field structure through measurements of dichroic polarization of point sources. 
\par
In $\S 2$ we describe the observations and data reduction. We derived the polarization maps of NGC 2024 and constrained the illuminating source of the large-scale IRNe through the polarization vector analysis in $\S 3.1$.  In $\S 3.2$ we present the newly found small-scale IRNe associated with young stars, and discuss their physical properties. In $\S 3.3$ we present the results of software aperture polarimetry of point sources. The polarization properties of point sources are investigated based on the measurements of both polarization degrees and colors due to dust extinction. In addition, we reveal the magnetic field structure over NGC 2024 obtained from the polarization angle of each source, and compare it with the data taken at other wavelengths. Our conclusions are summarized in $\S 4$. 

\section{OBSERVATIONS AND DATA REDUCTION}
We carried out near-infrared polarimetric observations of NGC 2024 using the imaging polarimeter SIRPOL (polarimetry mode of the SIRIUS camera: Kandori et al. 2006) on the 1.4 m telescope IRSF (Infrared Survey Facility) at the South African Astronomical Observatory (SAAO). SIRPOL consists of a single-beam polarimeter (rotating achromatic [$1-2.5 \mu $m] wave plate and high-extinction-ratio wiregrid analyzer) and the $JHK_{\rm s}$-simultaneous imaging camera SIRIUS (Nagayama 2003), which has three $1024 \times 1024$ HgCdTe infrared detectors (HAWAII array). IRSF/SIRPOL enables deep and wide-field ($7.7' \times 7.7'$ with a scale of $0.45''$ pixel${}^{-1}$) imaging polarimetry at $JHK_{\rm s}$ simultaneously. Polarizations of all the 2MASS-detected sources (15.8, 15.1, and 14.1 mag at $J$, $H$, and $K_{\rm s}$, ${\rm S}/ {\rm N} = 10$) can be measured within the accuracy of 1 \%. 
\par
Observations were made on the night of 2006 February 13. The 1.25 $\mu$m ($J$), 1.63 $\mu$m ($H$), and 2.14 $\mu$m ($K_{\rm s}$) imaging polarimetry data of NGC 2024 were obtained simultaneously. The 10 sec exposures at 4 wave plate angles (in the sequence of $0^{\circ}$, $45^{\circ}$, $22.5^{\circ}$, and $67.5^{\circ}$) were performed at 10 dithered positions (1 set). We repeated the same set 9 times toward the object. The sky frame were taken in the same manner. The total integration time toward NGC 2024 was 900 seconds per wave plate angle. The typical seeing during the observations was $\sim 1.4 ''$ (3 pixels) at $J$. We obtained twilight flat at the beginning and end of nights. In order to make good flat frames, we combined 23 sets of twilight flat observations (17 nights) taken in the period from December 2005 to February 2006. 
\par
We processed the observed data in the same manner as described in Kandori et al. (2006) using the IRAF\footnote{IRAF is distributed by the National Optical Astronomy Observatories, which are operated by the Association of Universities for Research in Astronomy, Inc., under cooperative agreement with the National Science Foundation.} (flat-field correction, median sky subtraction, and frame registration). 
The Stokes $I$, $Q$, and $U$ images are obtained using $I=(I_{0^{\circ}} + I_{45^{\circ}} + I_{22.5^{\circ}} +I_{67.5^{\circ}}) / 2$, $Q = I_{0^{\circ}} - I_{45^{\circ}}$, and $U = I_{22.5^{\circ}} - I_{67.5^{\circ}}$. The image of the polarization degree $P$ and polarization angle $\theta$ can be derived by $P = \sqrt{Q^2 + U^2} / I$ and $\theta = 0.5 \times {\rm atan}(U/Q)$. 
We calibrated $I$ sky level to be consistent with the surface brightness derived using 2MASS images covering a much larger field-of-view. 
The $10 \sigma$ limiting magnitudes for surface brightness of $I$ were 19.6, 19.3, and 17.3 mag arcsec${}^{-2}$ for $J$, $H$, and $K_{\rm s}$, respectively. In Fig.1 we show images of the $JHK_{\rm s}$-composite intensity ($I$) and polarized intensity ($PI$). We note that the PSF of unpolarized stars are not perfectly canceled on the $PI$ image, because the seeing size changes slightly during the observations to take $Q$ and $U$. 
\par
For the source detection and photometry on Stokes $I$ images, we used the IRAF {\it daophot} package (Stetson 1987). We detected stars having a peak intensity greater than $10 \sigma$ above local sky background. The overlooked sources in automatic detection were collected by eye inspection. The detected number of sources was 240, 365, and 376 at $J$, $H$, and $K_{\rm s}$, respectively. We measured instrumental magnitudes of stars with aperture photometry. The aperture radius was 3 pixels. The number of stars detected in all three bands with a photometric error of less than 0.1 mag is 211. The $10 \sigma$ limiting magnitudes were 18.5, 18.0, and 16.5 mag for $J$, $H$, and $K_{\rm s}$, respectively. 
In order to obtain plate solutions, we matched pixel coordinates of detected sources and celestial coordinates of their counterparts in the 2MASS Point-Source Catalog (Cutri et al. 2003),  and applied the IRAF {\it imcoords} package to the matched list. The coordinate transformation error is $\sim 0.1''$ (rms). 
We performed photometric calibration using the 2MASS-PSC. The magnitude and color of our photometry were transformed into the 2MASS system using, 
\begin{eqnarray}
{\rm MAG}_{\rm 2MASS} &=& {\rm MAG}_{\rm IRSF} + \alpha {}_{1} \times {\rm COLOR}_{\rm IRSF} + \beta {}_{1}, \\
{\rm COLOR}_{\rm 2MASS} &=& \alpha {}_{2} \times {\rm COLOR}_{\rm IRSF} + \beta {}_{2}, 
\end{eqnarray}
where the coefficients $\alpha {}_{1}$ are $0.029$, $-0.024$, and $0.008$ for $J$, $H$, and $K_{\rm s}$, respectively. The coefficients $\alpha_{2}$ for $J-H$ and $H-K_{\rm s}$ are $1.032$ and $0.997$, respectively. We note that the coefficients $\beta_{1}$ and $\beta_{2}$ include both zero point and aperture correction. 
\par
We carried out software aperture polarimetry of point sources on the combined intensity images for each wave plate angle ($I_{0^\circ}$, $I_{22.5^\circ}$, $I_{45^\circ}$, and $I_{67.5^\circ}$). This is because the center of point sources (i.e., aperture center) cannot be determined satisfactory on the $Q$ and $U$ images. We discuss the results in the $H$ band where nebula contamination is less than in the $J$ band and scattering efficiency is higher than in the $K_{\rm s}$ band. Based on aperture photometry data of each angle image, the Stokes parameters for each source were derived as described above. Point sources on each angle image have a slightly different PSF due to varying seeing size ($\sim 1.3'' - 1.5''$ at $J$). We thus used different apertures for each image in order to measure the same fraction of stellar flux falling in the aperture. The adopted aperture sizes were 3.00, 3.24, 3.08, and 3.27 pixels for $I_{0^{\circ}}$, $I_{22.5^{\circ}}$, $I_{45^{\circ}}$, and $I_{67.5^{\circ}}$, respectively. The sky annulus was set to 10 pixels with 5 pixel width. Since polarization degree $P$ is a positive quantity, derived $P$ values tend to be overestimated especially for low S/N sources. To correct the bias we calculated the debiased $P$ using $P_{\rm db} = \sqrt{P^{2} - {\delta P}^{2}}$, where $\delta P$ is the error in $P$ (Wardle \& Kronberg 1974). 

\section{RESULT AND DISCUSSION}
\subsection{Large-Scale Infrared Reflection Nebula (IRN)}
Although near-infrared ($H$ and $K$) polarimetry around a small nebula associated with FIR 4 was reported previously (Moore \& Yamashita 1995), no wide-field polarimetry of NGC 2024 has been carried out. Our wide and deep observations with SIRPOL revealed the distribution of near-infrared polarization in NGC 2024 for the first time. In Fig. 2, we show the $I$, $PI$, $P$, and ${\theta}$ of NGC 2024 in the $H$ band. 
The central dark lane running north-south is located foreground of the nebula and obscures the light from H II region (Fig. 2: upper-left). In the same panel, we show the location of far-infrared sources (FIR 1-7, Mezger et al. 1988, 1992) associated with a filament of dense material, $\lq \lq$star-forming ridge$"$,  located just behind the H II region. Among the far-infrared sources, FIR 4 is associated with a 2 $\mu$m source (Moore \& Chandler 1989) as well as reflection nebula (Moore \& Yamashita 1995), both of which are also confirmed with our data (see $\S 3.2$). In addition to the large scale IRN, there are smaller scale IRN associated with stars, which we discuss in the next section. 
We found prominent and extended {\it polarization} nebulosities over NGC 2024 for the first time (Fig. 2: upper-right). In contrast to the other regions, e.g., M 42, the exciting source of NGC 2024 is not directly identified due to the heavy obscuration by the central dust lane. Although IRS 2 was thought to be the exciting source of NGC 2024 until recently (e.g., Grasdalen 1974), Bik et al. (2003) proposed the the dominant ionizing source of NGC 2024 is IRS 2b located $5''$ north-west of IRS 2 (Fig. 2: upper-left). 
\par
Our imaging polarimetry data can be used to constrain the location of ionizing source(s) of the nebula through the polarization vector analysis assuming that the ionizing source and the illuminating source are the same. The vector map is shown in Fig. 3. We found the vector patterns clearly appear centrosymmetric, and normal lines of each vector intersect at the position near the center of NGC 2024 shown by white plus symbol. The location of dominant illuminating source(s) of NGC 2024 is then determined to be located within the white circle ($1 \sigma$ error circle) in Fig. 3. We note that the locations of both IRS 2 and IRS 2b are included in the error circle but IRS 2b is closer to the center. The vector patterns at $J$, $H$, and $K_{\rm s}$ are consistent with each other and result in the coincident symmetry center position. 
We colored polarization vectors black or green based on the angle $\theta$ between each vector and the line toward white plus symbol; vectors which seem to follow centrosymmetry, showing $\theta = 90 \pm 20$${}^{\circ}$, are in white, and other vectors are in green. Polarization vectors are mostly colored in white, indicating that most of the large-scale nebulosities are illuminated from star(s) in the central region enclosed by white circle. The green vectors tend to be distributed over the north-south dark dust lane and around small-scale IRNe, which can be explained by superposition of the large-scale nebulosities and extra polarization fields, i.e., dichroic absorption of foreground materials and intrinsic polarization of small-scale IRNe. 

\subsection{Small-Scale IRN}
We found five new small-scale IRNe (IRN 1-5) associated with stars that are shown in the $PI$ image in Fig. 2. An object enclosed by dashed line circle is the IRN associated with FIR 4, which was studied previously by Moore \& Yamashita (1995) with near-infrared polarimetry. We show $PI$ and vector maps of IRN in Fig. 4. It is clear that the pattern of the polarization vector appears centrosymmetric indicating that the nebulosity is associated with the central star, and responsible for the structure of circumstellar matter (e.g., disk/envelope systems) that produces strongly polarized light through dust scattering. 
The small nebulae in a larger scale nebulosity or embedded in the cloud should be treated with caution, because the polarization toward such nebulae is a superposition of different polarization vectors (i.e., scattered light polarization and dichroic polarization). We subtracted local background (constant value) around small-scale IRNe from the Stokes $Q$ and $U$ images in order to suppress the contribution from larger nebulosities. We determined the $Q$ and $U$ background levels at the position 20$''$ away from the center of IRN, and simply subtracted it from the original. We did not take the effect of dichroic polarization into account, because the resulting vector pattern of IRNe after the subtraction of local background was reasonably centrosymmetric and the estimate of dichroic polarization is not straightforward. 
\par
All IRNe have certain structures on the $PI$ images. The spatial extent of small-scale IRNe is typically about $5000-10000$ AU on the $PI$ images, comparable to the size of $\lq \lq$envelopes$"$ rather $\lq \lq$disks$"$. 
IRNe show an elongated polarized emission pattern whose position angle is roughly 90, 135, 30, 45, and 135${}^{\circ}$ for IRN 1-5, respectively. The emission pattern of IRN 1, 2, and 4 is clearly butterfly-shaped, and the dark lane extending perpendicular to the bright region can be seen in these IRNe. For IRN 3 and 5, the elongated emission pattern can only be seen on one side of the central star. 
These characteristics agree with the polarization picture of disk/envelope system around young stars (e.g., Nagata, Sato, \& Kobayashi 1983; Hodapp 1984; Sato et al. 1985). The radiation from the star causes heavy extinction toward the direction of disk (seen as a dark lane), but can escape toward the polar directions and then scattered by dust at the surface of the envelope or the cavity created by the outflow (seen as a butterfly-shaped or monopolar IRN). 
\par
In Fig. 5, we show $PI$ and vector maps of the IRN associated with FIR 4 in the $J$, $H$, and $K_{\rm s}$ bands. The counterpart of FIR 4 is best seen in the $K_{\rm s}$ band and located at ($\Delta$R.A., $\Delta$Dec.) = ($7.2''$, $-7.2''$) in Fig. 5. Our vector maps show good agreement with those in the $H$ and $K$ bands reported by Moore \& Yamashita (1995). 
The IRN appears as a monopolar shape with position angle of $\sim 135^{\circ}$, and shows a centrosymmetric vector pattern centered on FIR 4. The axis of IRN is consistent with that of the molecular outflow associated with FIR 4 (Chandler \& Carlstrom 1996). The position angle of the polarization vector toward FIR 4 is $\sim 45^{\circ}$, perpendicular to the axis of IRN, which is the general tendency in young stellar objects with IRN (see above). 
\par
In Table 1, we summarize the observational properties of IRNe and its central stars. The polarization degree of each IRN (column 4-6th) is not large, because unpolarized light from the HII region (e.g., free-free ionized gas emission) dilute the polarized light. If we subtract the background intensity around IRN, the polarization degrees increase to the level comparable with typical reflection nebulae in other star forming regions as shown in parentheses. The IRN 1-5 are associated with stars IRS 12, 4, 3, 28 and 30, respectively. From their near-infrared colors, we found that the central stars of five IRNe are young, especially in a pre-main-sequence phase. 
The infrared color of the five sources were investigated in the $JHKLN$ bands by Haisch et al. (2001). Based on the $2.2 - 10$ $\mu$m spectral index $\alpha$ ($ = d \log \lambda F_{\lambda} / d \log \lambda$) measurements, the five sources, IRS 12 (IRN 1), IRS 4 (IRN 2), IRS 3 (IRN 3), IRS28 (IRN 4), and IRS 30 (IRN 5), were classified into Class II, Class I, Class II, Class II, and Flat Spectrum, respectively. 
This is consistent with the fact that they are accompanied by nebulous circumstellar materials. Following the method described in $\S \S$ 2, we performed aperture polarimetry of the central stars (column 12-13). Except for the saturated source IRS 12 and the source IRS 3 with faint IRN, we found that the polarization vector angle $\theta$ of stars is roughly perpendicular to the major axis of IRN (column 3). The characteristic is in good agreement with those for young stellar objects accompanied by IRNe. These sources serve as good target for future studies in higher resolution at various wavelengths (e.g., large Opt-IR telescopes, ALMA) as well as in theoretical modeling. 

\subsection{Aperture Polarimetry of Point-like Sources} 
We performed software aperture polarimetry of point-like sources in order to measure the integrated polarization of possible unresolved nebulae. Our photometry and polarimetry of 211 stars with $JHK_{\rm s}$ detections are shown in Table 2. The polarization vectors in the $J$, $H$, and $K_{\rm s}$ bands are quite consistent with each other. 
For the sources with $\delta {\theta}_{\lambda}$ of less than 3${}^{\circ}$ (63 sources), the correlation coefficient for (${\theta}_{H}$, ${\theta}_{J}$) and (${\theta}_{H}$, ${\theta}_{K_{\rm s}}$) is 0.99 and 0.98, respectively. The mean of the difference in ${\theta}_{\lambda}$ is small, which is ${-1.4}^{\circ}$ for $<{\theta}_{H} - {\theta}_{J}>$ and ${-1.7}^{\circ}$ for $<{\theta}_{H} - {\theta}_{K_{\rm s}}>$. 
For the sources with ${P}_{\lambda} / \delta {P}_{\lambda}$ of greater than 10 (64 sources), the linearly fitted slope of ${P}_{H}$ vs. ${P}_{J}$ and ${P}_{H}$ vs. ${P}_{K_{\rm s}}$ diagram is 1.52 and 0.50, with the correlation coefficient of 0.98 and 0.91, respectively. These slopes are consistent with the empirical value of 1.61 and 0.61 from the relation $P \propto {\lambda}^{- \alpha}$, where $\alpha = 1.8 \pm 0.2$ (Whittet 1992). 
In the following, we discuss the aperture polarimetry results in the $H$ band where nebula contamination is less than in the $J$ band and scattering efficiency is higher than in the $K_{\rm s}$ band. 

\subsubsection{Highly Polarized Sources (HPS)}
We can detect candidate stars with intrinsic polarization (i.e., with circumstellar material) by selecting the sources with larger polarizations than values estimated from the dichroic absorption due to foreground material. Firstly, we made two color diagram for all the sources detected in the $JHK_{\rm s}$ bands in order to classify the sample into three groups of the dwarf+giant, PMS stars, and protostars. The result is shown in Fig. 6a. We found 143 stars in the dwarf+giant star region (filled circle), 63 stars in the PMS star region (open diamond), and 5 stars located on right to the PMS region, i.e., protostar region (open circle). 
We plotted these sources in $H-K_{\rm s} - 0.15$ versus $P_{H}$ diagram shown in Fig. 6b. The value of 0.15 mag is the mean intrinsic color of stars (mostly G to M dwarfs) toward the object calculated using the Galaxy model (Wainscoat \& Cohen, 1992; Cohen, 1994), so that the horizontal axis of Fig. 6b corresponds to the mean $E_{H-K_{\rm s}}$. 
The dashed line denotes the mean relationship between extinction and polarization for interstellar molecular clouds, $P_{K} {\rm (\%)} = 2.23 \tau _{K}^{3/4}$, from Jones (1989). 
A linear least absolute deviation fitting to our $H-K_{\rm s} - 0.15$ and $P_{H}$ data resulted in $P_{H} = 4.21 \times E_{H-K_{\rm s}} = 1.69 \times A_{H}$ (solid line), where we forced the intercept value to be zero in the fitting. Our fitting result shows a similar relationship to the relation from Jones (1989). The polarization of stars with $P_{H}/A_{H} \sim 1.69$ is dominated by dichroic polarization produced by aligned interstellar dust grains. Observational upper limit of the relationship (dot-dashed line) can be described using the equation $P_{\rm max} = {\rm tanh} (\tau _{K})$ where $\tau _{K} = (1-\eta)\tau _{K} / (1+\eta)$ when the parameter $\eta$ is set to 0.875 (Jones 1989). 
\par 
The sources with larger $P_{H}/A_{H}$ values can be better candidates to show excess (intrinsic) polarizations, and hense good targets for further observations in order to study circumstellar materials around stars. We calculated $P_{H}/A_{H}$ for each star, where $A_{H}$ was measured on the two color diagram (Fig. 6a). The histogram of $P_{H}/A_{H}$ is shown in Fig. 7. In the histogram, there is a clear peak around $P_{H}/A_{H} \sim 1.69$, which corresponds to the sources mainly affected by interstellar polarization. 
We found that the dwarf+giant stars (filled circle) and PMS stars (open diamond) show similar distribution in Fig. 6b. If we regard the stars with $P_{H}/A_{H} \ge 3.38$, two times larger $P_{H}/A_{H}$ than the interstellar polarization, as highly polarized sources (HPS), the fraction of HPS is 32 \%,  31 \%, and 35 \%  for all the stars, dwarf+giant stars, and, PMS stars, respectively. The protostar candidates are not included in this statistics, because their intrinsic colors are not known. It is an interesting result that there is no tendency of higher polarization in earlier evolutionary stages or larger infrared excess. However, since the part of the locus of classical T Tauri stars overlaps with the reddening band, some fraction of PMS stars can be included in the sample of dwarf+giant stars. Thus the HPS fraction for the dwarf+giant group may cause significant contamination from PMS stars and caution must be applied. 

\subsubsection{Stellar Mass versus $P_{H}/A_{H}$}
It is of particular interest to investigate the fraction of HPS against the luminosity of stars ($\propto$ mass), because the information should be closely related to the dependence of the population of stars with circumstellar disk on the mass of stars. 
Though the probability of disk inclination, orientation, and suffered extinction from foreground materials is expected to be random, other factors, the shape (scale height) and evolutional timescale (lifetime) of disk, may have some dependence on stellar mass. 
\par
It is likely that the effect of disk lifetime on the detection ratio of HPS is not significant. As shown in Fig. 6a, many sources are located in the Class II region in the two color diagram. By using $J-H$ vs. $K-L$ diagram, Haisch et al. (2000) reported that infrared excess stars (Class II and protostars) occupy $\sim 80$ \% fraction of the stellar population in this region. Since the age of most sources could be younger than or comparable with disk lifetime ($\sim 10$ Myr, Strom \& Edwards 1993), the observed HPS fraction should not be affected by the disk lifetime. 
\par
Theoretically very low mass stars (i.e., brown dwarfs) are expected to have a more flared (larger scale height) disk than those for more massive stars due to its small gravity (Walker et al. 2004). Highly flared disks result in a large fraction of obscured stars due to extinction, so that only scattered light is observed, making the source a HPS. Considering the disk optical depth (i.e., size and mass) and its flaring for several models, the disk obscuring probability of brown dwarfs in the $K$ band ($2.2 \mu$m) was estimated to be larger than that for Class II stars by a factor of $\sim 1-3$ (Walker et al. 2004). Thus, a stellar mass dependence of HPS detection rate may be observed. 
\par
To derive intrinsic luminosities, we performed dereddening of stars by subtracting $A_{H}$ from observed $H$ magnitudes. We then derived the fraction of sources with $P_{H} / A_{H} > 3.38$ against dereddened $H$ mag at each bin separated by 1 mag. The result is shown in Table 3. Note that stars with $H \ge 13.3$ mag are brown dwarfs at the employed distance of 415 pc (Anthony-Twarog 1982). We found that the distribution of polarization fraction remains constant from low (brown dwarfs) to higher luminosity (solar-type) stars, although the error in polarization fraction is not small. According to the argument above, our finding suggests that the dependence of disk scale height on the mass of central star is not significant in spite of the theoretical model that predicts more flared disks for lower mass stars. It is noteworthy that a very similar result is obtained in the OMC-1 region (Kusakabe et al. 2006). Quantitative studies on the origin of flat HPS distributions will be discussed in a later paper. 
\par
Recently, Levine et al. (2004) reported the spectroscopic studies of low-mass stars and brown dwarfs distributed over $10' \times 10'$ region in NGC 2024. Their list of 71 stars with known spectral type can be more useful to investigate the relationship between stellar mass and $P_{H}/A_{H}$. We matched the coordinates of their sources with our polarimetry, and found 49 pairs. We show the stellar mass versus $P_{H}/A_{H}$ diagram in Fig. 8. The vertical and horizontal broken lines denote the upper limit of brown dwarf mass (0.08 $M_{\odot}$) and $P_{H}/A_{H} = 3.38$, respectively. It is clear that there is no obvious tendency for higher $P_{H}/A_{H}$ for lower mass stars. If we divide the sources into two groups at the line of $ M = 0.08 M_{\odot}$, the median $P_{H}/A_{H}$ value with ${\rm (rms)} / \sqrt{\rm n}$ error is $2.11 \pm 0.53$ and $1.62 \pm 0.43$ for the lower and higher mass groups, respectively. 
There can be a tendency of slightly higher $P_{H}/A_{H}$ for brown dwarfs than that for solar-type stars, but, within 1$\sigma$, those two values are identical. We should also note the difference in suffered mean extinction for both groups is small ($<A_{H}>_{\rm brown\ dwarf}=1.5$ mag and $<A_{H}>_{\rm soalr-type}=1.9$ mag). We conclude that the stellar mass vs. $P_{H}/A_{H}$ relationship is generally flat in the mass range from brown dwarfs to solar-type stars. 
\par
In Fig. 8, the sources located above the horizontal broken line can be classified as HPS. Their properties are summarized in Table. 4\footnote{Full listing of all sources is shown in Table 5 (electric version only).}. We found ten HPSs which consist of five brown dwarfs and five low-mass stars. It is most likely that these sources are associated with circumstellar materials which produce highly polarized light through dust scattering. The existence of circumstellar disk around young brown dwarfs has been indirectly suggested (e.g., infrared excess emission, Muench et al. 2001). In addition to a few samples revealed by imaging of disks using HST (Bally et al. 2000) and outflow detection with spectro-astrometry (Whelan et al. 2005), our five samples serve as $\lq \lq$direct$"$ evidence for the existence of disk/envelope system around brown dwarfs (see also, Tamura et al. 2006; Kusakabe et al. 2006). 

\subsubsection{Magnetic Field Structure}
The aperture polarimetry of stars provides important information on the direction of magnetic fields. If we assume the normal grain alignment, i.e., the spin axis of elongated dust grain align parallel to the magnetic fields (Davis \& Greenstein 1951), the direction of magnetic fields projected onto the sky can be inferred from the direction of polarization vectors of stars ($B \parallel E$, e.g., Weintraub, Goodman, \& Akeson 2000). Though a grain alignment mechanism caused by gas streaming motion (Gold 1952) may be the dominant process near or around sources with molecular outflows, we believe our assumption is plausible, because we discuss the structure of magnetic fields not only around each outflow sources but also across the entire surface of NGC 2024. 
\par
In Fig. 9, we show the histogram of polarization position angle, $\theta _{H}$, of stars in NGC 2024. It is clear that the stars with $P_H \ge 3$ \% show a peak at the position angle of $110^{\circ}$. The large scale distribution of magnetic field in NGC 2024 is thus expected to be in the direction of $110^{\circ}$. In the right panel, the stars with low polarization ($P_H < 3$ \%) shows less clear tendency. In the histogram, we show the number of protostars, PMS stars, and dwarf+giant stars in dark gray, gray and white colors, respectively. There is no large difference in the shape of histograms among three groups with different evolutionary stage. 
\par
We show the distribution of polarization vectors for each star in Fig. 10. Though the overall distribution of polarization angle is $\sim 110^{\circ}$, we found a bend of magnetic fields around the northwest of the star-forming ridge of NGC 2024. Since the bend of magnetic field orientation is localized on the polarization map, it is most likely that the structure is not associated with ambient material around NGC 2024 but associated with a smaller-scale structure inside the region, i.e., the star forming ridge. 
\par
Matthews, Fiege, \& Moriarty-Schieven (2002) studied the magnetic field structure of NGC 2024 based on the sub-mm dust continuum observations using the polarimetry mode of JCMT/SCUBA. We overlaid the polarization vector of stars on their polarization map in Fig. 11. Vectors from the sub-mm emission polarimetry (blue lines) were rotated by 90 degrees so as to be compared with the $E$-vector of dichroic polarization in the $H$ band (yellow lines). We futher plotted rotated (90 degrees) polarization vectors taken with 100 $\mu$m dust emission (red lines, Hildebrand et al. 1995; Dotson et al. 2000). 
The background image in Fig. 11 is the 850 $\mu$m dust continuum intensity map taken with JCMT/SCUBA. The gray broken line shows the contour of 0.9 Jy beam${}^{-1}$, which corresponds to $A_{V} \sim 50 $ mag when we use the dust opacity $\kappa {}_{850\ \mu {\rm m}} = 0.02$ cm$^2$ g$^{-1}$ and dust temperature $T_{\rm d} = 18$ K. The region enclosed by the gray broken line is too highly obscured to be traced at the near-infrared wavelengths. 
\par
We found that the direction of magnetic field obtained from both dust emission (blue and red lines) and extinction (yellow lines) is generally consistent with each other in the region around and outside the contour of $A_{V} \sim 50$ mag. The agreement between dichroic and dust emission polarizations suggests that dichroic polarization traces well the magnetic field structure if lying dense dusty material is less than or as much as its penetration depth. 
On the other hand, dichroic polarization vectors do not show good correlation with those obtained from dust emission inside the contour of $A_{V} \sim 50$ mag. It is a natural result if we consider that the dichroic polarization in the $H$ band only trace the magnetic field foreground to the star-forming ridge. The disagreement between dichroic and dust emission polarizations suggests that magnetic field structure toward the star-forming ridge traced using the dust emission polarimetry is localized in both the plane-of-sky and the line-of-sight direction. 
We note that the line-of-sight variation in magnetic field strength ($B_{\parallel}$) toward the star-forming ridge is also observed by Crutcher et al. (1999) based on the measurements of the Zeeman splitting of the absorption lines of HI and OH with Very Large Array (VLA). 
On the basis of the sub-mm dust emission polarimetry data, Matthews, Fiege, \& Moriarty-Schieven (2002) explained the magnetic field structure toward the star-forming ridge using two different models; one is helical magnetic fields surrounding a curved filamentary cloud, and another is magnetic fields swept by the ionization front of the expanding HII region. It is hard to differentiate the two models based on our near-infrared dichroic polarization data, because the data can not trace the dense region ($A_{V} \simgt 50$ mag) of the star-forming ridge. 
\par
It is noteworthy that our result is contrary to the view that the polarization of background starlight in the optical to near-infrared does not trace magnetic fields (i.e., grain alignment) inside cold and dense molecular clouds (e.g., Goodman et al. 1995). For example, Arce et al. (1998) suggested that the dichroic polarization of background stars toward the Taurus dark cloud only trace the region up to $A_{V} \sim 1.3$ mag, indicating that the dichroic polarization does not trace magnetic field inside cold dark clouds due to significant depolarization effect. 
Their criticism may not be the case for other star forming regions where the environment is different 
(e.g., higher temperature\footnote{The temperature of gas and dust was measured as $T_{\rm kin} \simlt 20$ K (Gaume, Johnston, \& Wilson 1992; Mauersberger et al. 1992) and $T_{\rm d} \sim 16 - 18$ K (16 K toward the star-forming ridge: Mezger et al. 1988, 18 K toward the Orion B South region: Johnstone et al. 2006) in NGC 2024.}) from those in cold dark clouds. 
In addition to the NGC 2024 region, a good consistency in magnetic field structure obtained using dichroic polarization and dust continuum polarization is confirmed in the OMC-1 (Houde et al. 2004; Kusakabe et al. 2006) and NGC 2071 (Matthews, Fiege, \& Moriarty-Schieven 2002; Tamura et al. 2006) region. We thus conclude that the dichroic polarization at near-infrared wavelengths is a good tracer of magnetic field structures than previously thought and wide-field near-infrared polarimetry is useful to reveal the magnetic field structures in various star-forming regioins. 

\section{SUMMARY}
We conducted deep and wide-field $JHK_{\rm s}$ imaging polarimetry toward NGC 2024, a massive star-forming region in the Orion B cloud. This is the first imaging polarimetry covering almost all of the HII region as well as embedded young cluster in NGC 2024. Our conclusions are summarized as follows. 
\begin{enumerate}
\item We found a prominent and extended infrared reflection nebula (IRN) over NGC 2024 on our polarization images. We constrained the location of illuminating source of the nebula through the analysis of polarization vectors. A massive star, IRS 2b with a spectral type of O8 $-$ B2, is located in the center of the symmetric vector pattern. 
\item We discovered five small-scale IRNe associated with YSOs in our polarization images. The pattern of polarization vector around the stars appears centrosymmetric indicating that these IRNe are illuminated by each central star and are responsible for the structure of circumstellar material that produces strongly polarized light through dust scattering. The polarized intensity of each IRN shows a butterfly-shaped emission pattern extending over $5000 - 10000$ AU, which agrees well with the picture of disk/envelope system around young stars. 
\item We performed software aperture polarimetry of the 211 point-like sources detected on the Stokes $I$ images at $JHK_{\rm s}$. We cataloged 64 highly polarized sources (HPS) as candidates associated with circumstellar material by selecting the sources with stronger polarizations than estimated from dichroic extinction. 
\item We investigated the fraction of highly polarized sources against the intrinsic luminosity of stars ($\propto$ mass), and found that the source detection rate remains constant from low (brown dwarfs) to higher luminosity (solar-type) stars. Since most stars in the region are young enough to have circumstellar disks, possible differences in disk lifetime should not affect our statistics. Though a stellar-mass-dependence of disk scale height (flaring) is a possible factor to naturally change the detection rate of highly polarized source, our result indicates that the effect is not significant in spite of the theoretical prediction of more flared disks for lower mass stars. We confirmed the result using our polarimetry of the stars with known spectral types toward NGC 2024.. 
\item By comparing our polarimetry with the spectro-photometry by Levine et al. (2004), we found five brown dwarfs with highly polarized integrated emission. These sources serve as direct evidence of circumstellar material around brown dwarfs. 
\item We investigated the magnetic field structure of NGC 2024 through the measurements of dichroic polarization of point sources. The position angle of projected magnetic field across the region is generally found to be 110${}^{\circ}$. We also found a bending of magnetic fields around the northwest of the star-forming ridge of NGC 2024. 
\item In the region of $A_{V} \le 50$ mag, we found good consistency in magnetic field structures obtained using dichroic polarization at near-infrared and dust continuum polarization at sub-mm/far-infrared wavelengths. Similar results are also found in the OMC-1 and NGC 2071 regions. These results strongly indicate that the dichroic polarization at near-infrared wavelengths traces magnetic field structures inside dense molecular clouds. 
\item In the region of $A_{V} > 50$ mag over the star-forming ridge, the dichroic polarization does not show good correlation with those obtained from dust emission. It is a natural result if we consider that the dichroic polarization at near-infrared wavelengths can only trace the alignment of foreground dust. The disagreement between dichroic and dust emission polarizations suggests that magnetic field structure toward the star-forming ridge traced using the dust emission polarimetry is localized in both the plane-of-sky and the line-of-sight direction. 
\end{enumerate}

\bigskip 
 
We are grateful to Shuji Sato and Phil Lucas for there helpful comments and suggestions. Thanks are due to the staff in SAAO for their kind help during the observations. We thank the referee, Munetaka Ueno, for useful comments on the manuscript. We also thank Doug Johnstone for providing his JCMT/SCUBA data of NGC 2024. The IRSF/SIRIUS project was initiated and supported by Nagoya University, National Astronomical Observatory of Japan, and University of Tokyo in collaboration with South African Astronomical Observatory under a financial support of Grants-in-Aid for Scientific Research on Priority Area (A) No. 10147207 and No. 10147214, and Grants-in-Aid No. 13573001 and No. 16340061 of the Ministry of Education, Culture, Sports, Science, and Technology of Japan. M. T. and R.K. acknowledges support by the Grants-in-Aid (No. 16077101, 16077204, 16340061). J. H. acknowledges support by PPARC.

\clearpage
\onecolumn

\begin{landscape}
\begin{center}
\smallskip
Table 1: Properties of small-scale IRNe
\smallskip
{\small
\begin{tabular}{cccccccccccccccc} \hline \hline
              \multicolumn{7}{c}{Nebulae} &  & \multicolumn{7}{c}{Central Stars}    & \\
		   \cline{1-6}   \cline{8-16}
      ID        &   Size$^{a}$   &  PA$^{a}$  & $P_{J{\rm , max}}^{b}$  & $P_{H{\rm , max}}^{b}$ & $P_{K_{\rm s, max}}^{b}$   &  &
		No.$^c$	   &     Name$^d$  &  $J^e$  & $H^e$  & $K_{\rm s}^e$  & Class$^f$ &  $P_{H}$  &  ${\theta}_{H}$  \\
		  	  &    (AU)  &  ($^{\circ}$)      & (\%)        &    (\%)      &   (\%)          &   &
              &                &  (mag)        & (mag)           & (mag)  &   &   (\%)   &   (${}^{\circ}$) \\
\hline
IRN1      &   10000 & 90        &    12.5(19.0) & 9.0(15.0)    &   ---   &         &  ---    & IRS 12 &  10.36 & 9.33 & 8.63   & II  &  ---         &  ---         \\ 
IRN2      &   7000 & 135        &    16.5(24.0) & 14.5(21.0)  & 8.0(10.0)  &   &  138 & IRS 4   &  12.45 & 10.49 & 9.19  & I &  $8.1 \pm 0.11$ &  $57 \pm 0.4$   \\ 

IRN3      &   6000  & 30         &      ---      &   13.0(22.0)  &  9.0(12.0) &  & 148  & IRS 3  &  13.66  &  11.34  &  9.81  & II &$7.0 \pm 0.10$ & $104 \pm 0.4$  \\

IRN4      &   5000 & 45          &    28.0(30.0) & 27.0(29.0)  & ---           &   &  191    & IRS 28  &  13.95 & 11.85 & 10.62 & II & $5.1 \pm 0.06$ &  $127 \pm 0.4$  \\ 
IRN5      &   6000 & 135        &    ---             & 28.0(28.0)  & ---           &  &  194    & IRS 30 &   12.59 & 10.48 & 9.32  & Flat Spectrum & $1.5 \pm 0.02$  &  $85 \pm 0.3$  \\
\hline
\end{tabular}
}
\begin{list}{}{}
\item[$^a$] {\small Size and position angle of IRNe on the $PI$ images. The size of the nebula having intensity greater than 10$\sigma$ above background was measured as the size of the IRN. The position angle denotes the orientation of polarized nebulosity of each IRN.} 
\item[$^b$] {\small Maximum polarization degree of IRNe was measured after the subtraction of local background on $Q$ and $U$ images. If we further remove the contribution from large-scale nebula by subtracting the local background intensity ($I$), the polarization degree of IRNe increase as shown in parentheses.} 
\item[$^c$] {\small Source number in Table 2.} 
\item[$^d$] {\small Star-ID based on Barnes et al. (1989).} 
\item[$^e$] {\small Magnitudes taken from 2MASS Point Source Catalog (Cutri et al. 2003).} 
\item[$^f$] {\small Evolutionary state of the central stars (YSOs) based on the $2.2 - 10$ $\mu$m spectral index measurements (Haisch et al. 2001).} 
\end{list}{}{}
\end{center}
\end{landscape}

\setcounter{table}{1}
\clearpage
\begin{center}
\smallskip
Table 2: Polarizations of Point-like Sources in NGC 2024
\smallskip
{\scriptsize
\begin{tabular}{ccccccccccc} \hline \hline
No. & HPS ID$^{a}$ & R.A.$^{b}$  & Dec.$^{b}$        &  $J$ & $H$ & $K_{\rm s}$ & $P_{H}$ & $A_{H}$$^{c}$ & $P_{H}/A_{H}$$^{d}$  &  $\theta _{H}$ \\
      &              & (${}^{\rm h}\ {}^{\rm m}\ {}^{\rm s}$) & (${}^{\circ} \ ' \ ''$) & (mag) & (mag) & (mag) & ($\%$) & (mag) & 
	  ($\%$ mag${}^{-1}$)  &  (${}^{\circ}$)  \\
\hline
  1 & --- & 5 41 47.25 & -1 51 08.3 & 17.64 & 15.84 & 14.75 &  3.25$\pm$1.62 & 1.24 &  2.63$\pm$1.31 &   5$\pm$12.8 \\
  2 & --- & 5 41 46.95 & -1 51 09.9 & 15.48 & 13.79 & 12.98 &  3.50$\pm$0.30 & 1.67 &  2.09$\pm$0.18 &  19$\pm$2.4 \\
  3 & --- & 5 41 44.65 & -1 51 12.5 & 16.87 & 14.50 & 13.26 &  5.47$\pm$0.70 & 2.71 &  2.02$\pm$0.26 &  50$\pm$3.6 \\
  4 & --- & 5 41 58.52 & -1 51 12.9 & 14.84 & 13.09 & 12.28 &  2.06$\pm$0.09 & 1.64 &  1.25$\pm$0.05 &  87$\pm$1.2 \\
  5 & HPS1 & 5 41 50.10 & -1 51 18.1 & 16.01 & 14.84 & 14.03 &  1.56$\pm$0.43 & 0.33 &  4.75$\pm$1.32 &  84$\pm$7.7 \\
  6 & --- & 5 41 45.64 & -1 51 22.8 & 14.74 & 12.74 & 11.68 &  2.44$\pm$0.11 & 2.23 &  1.09$\pm$0.05 &  74$\pm$1.3 \\
  7 & --- & 5 41 57.86 & -1 51 27.9 & 13.52 & 12.31 & 11.80 &  0.30$\pm$0.04 & 0.90 &  0.34$\pm$0.05 &  83$\pm$4.1 \\
  8 & --- & 5 41 48.31 & -1 51 30.5 & 17.50 & 15.53 & 14.39 &  2.82$\pm$1.51 & 1.51 &  1.86$\pm$1.00 & 164$\pm$13.5 \\
  9 & --- & 5 41 43.68 & -1 51 35.3 & 15.28 & 12.78 & 11.61 &  1.32$\pm$0.10 & 2.48 &  0.53$\pm$0.04 &  33$\pm$2.1 \\
 10 & --- & 5 41 43.89 & -1 51 39.8 & 17.79 & 14.27 & 12.57 &  1.58$\pm$0.35 & 3.78 &  0.42$\pm$0.09 &  50$\pm$6.2 \\
 11 & --- & 5 41 37.30 & -1 51 40.3 & 15.56 & 12.84 & 11.35 &  3.24$\pm$0.26 & 3.27 &  0.99$\pm$0.08 & 152$\pm$2.3 \\
 12 & --- & 5 41 45.09 & -1 51 44.3 & 14.21 & 11.60 & 10.29 &  4.25$\pm$0.05 & 2.93 &  1.45$\pm$0.02 &  44$\pm$0.3 \\
 13 & --- & 5 41 39.02 & -1 51 45.3 & 15.06 & 12.81 & 11.73 &  4.16$\pm$0.14 & 2.38 &  1.75$\pm$0.06 & 141$\pm$0.9 \\
 14 & --- & 5 41 41.77 & -1 51 45.6 & 13.93 & 12.35 & 11.63 &  1.31$\pm$0.07 & 1.41 &  0.93$\pm$0.05 & 164$\pm$1.4 \\
 15 & --- & 5 41 54.73 & -1 51 46.2 & 17.68 & 16.04 & 15.02 &  2.39$\pm$1.37 & 1.02 &  2.33$\pm$1.34 & 138$\pm$14.2 \\
 16 & --- & 5 41 46.52 & -1 51 48.8 & 17.66 & 15.29 & 13.84 &  3.80$\pm$0.87 & 1.95 &  1.95$\pm$0.45 &  57$\pm$6.4 \\
 17 & --- & 5 41 38.08 & -1 51 53.6 & 18.04 & 15.91 & 14.65 &  5.62$\pm$1.55 & 1.70 &  3.31$\pm$0.91 & 169$\pm$7.6 \\
 18 & --- & 5 41 45.38 & -1 51 56.6 & 17.91 & 13.81 & 11.42 & 11.96$\pm$0.25 & 4.23 &  2.83$\pm$0.06 &  28$\pm$0.6 \\
 19 & --- & 5 41 41.03 & -1 51 57.7 & 17.34 & 13.79 & 12.14 &  4.26$\pm$0.35 & 3.69 &  1.15$\pm$0.09 &  26$\pm$2.3 \\
 20 & --- & 5 41 37.26 & -1 51 57.9 & 17.20 & 14.88 & 13.67 &  4.62$\pm$0.37 & 2.62 &  1.76$\pm$0.14 & 161$\pm$2.3 \\
 21 & --- & 5 41 51.70 & -1 52 08.2 & 15.16 & 13.14 & 11.88 &  2.49$\pm$0.13 & 1.49 &  1.67$\pm$0.08 &  98$\pm$1.4 \\
 22 & --- & 5 41 35.37 & -1 52 11.5 & 14.83 & 12.70 & 11.40 &  3.12$\pm$0.16 & 1.65 &  1.89$\pm$0.10 & 144$\pm$1.5 \\
 23 & HPS2 & 5 41 32.83 & -1 52 12.5 & 14.29 & 12.65 & 11.65 &  7.04$\pm$0.07 & 1.02 &  6.88$\pm$0.06 & 131$\pm$0.3 \\
 24 & --- & 5 41 45.77 & -1 52 13.0 & 16.46 & 13.91 & 12.37 &  3.05$\pm$0.53 & 2.17 &  1.40$\pm$0.24 &  63$\pm$4.9 \\
 25 & --- & 5 41 34.52 & -1 52 13.4 & 15.39 & 14.12 & 13.40 &  2.91$\pm$0.62 & 1.37 &  2.11$\pm$0.45 & 145$\pm$5.9 \\
 26 & --- & 5 41 51.61 & -1 52 16.7 & 14.48 & 12.91 & 12.17 &  0.60$\pm$0.13 & 1.49 &  0.40$\pm$0.08 & 131$\pm$5.9 \\
 27 & --- & 5 41 55.68 & -1 52 18.6 & 12.77 & 10.29 &  9.17 &  1.46$\pm$0.01 & 2.25 &  0.65$\pm$0.01 & 144$\pm$0.3 \\
 28 & HPS3 & 5 41 31.86 & -1 52 19.9 & 12.31 & 10.97 & 10.45 &  2.56$\pm$0.02 & 0.72 &  3.56$\pm$0.03 & 127$\pm$0.2 \\
 29 & --- & 5 41 48.06 & -1 52 19.9 & 15.90 & 14.14 & 13.24 &  2.91$\pm$0.29 & 1.84 &  1.58$\pm$0.16 &  46$\pm$2.9 \\
 30 & --- & 5 41 40.27 & -1 52 20.5 & 15.40 & 12.66 & 11.26 &  2.18$\pm$0.11 & 3.14 &  0.70$\pm$0.03 & 159$\pm$1.4 \\
 31 & --- & 5 41 40.52 & -1 52 21.2 & 16.52 & 13.49 & 11.85 &  2.22$\pm$0.25 & 3.68 &  0.60$\pm$0.07 & 165$\pm$3.2 \\
 32 & --- & 5 41 50.90 & -1 52 23.6 & 15.63 & 14.34 & 13.55 &  0.85$\pm$0.31 & 0.59 &  1.44$\pm$0.52 &  44$\pm$9.7 \\
 33 & --- & 5 41 35.46 & -1 52 28.8 & 13.27 & 11.51 & 10.51 &  3.02$\pm$0.03 & 2.06 &  1.47$\pm$0.01 & 150$\pm$0.3 \\
 34 & --- & 5 41 48.94 & -1 52 29.7 & 16.12 & 13.77 & 12.60 &  4.09$\pm$0.19 & 2.59 &  1.58$\pm$0.07 &  72$\pm$1.4 \\
 35 & --- & 5 41 37.49 & -1 52 31.0 & 16.32 & 14.75 & 13.81 &  3.21$\pm$1.46 & 0.96 &  3.33$\pm$1.52 & 161$\pm$11.8 \\
 36 & --- & 5 41 31.62 & -1 52 31.6 & 14.32 & 13.12 & 12.52 &  1.70$\pm$0.13 & 1.09 &  1.57$\pm$0.12 & 122$\pm$2.2 \\
 37 & --- & 5 41 36.94 & -1 52 33.2 & 13.16 & 11.64 & 10.93 &  2.36$\pm$0.06 & 1.43 &  1.65$\pm$0.04 & 154$\pm$0.7 \\
 38 & --- & 5 41 36.86 & -1 52 37.5 & 14.16 & 12.74 & 11.96 &  2.02$\pm$0.20 & 1.48 &  1.36$\pm$0.14 & 165$\pm$2.9 \\
 39 & --- & 5 41 51.40 & -1 52 41.0 & 17.01 & 15.03 & 13.76 &  2.30$\pm$0.57 & 1.38 &  1.66$\pm$0.41 &  70$\pm$6.9 \\
 40 & --- & 5 41 36.70 & -1 52 42.9 & 13.24 & 11.94 & 11.40 &  2.07$\pm$0.11 & 0.98 &  2.11$\pm$0.11 & 169$\pm$1.5 \\
 41 & --- & 5 41 48.08 & -1 52 43.0 & 18.06 & 15.80 & 14.52 &  1.32$\pm$1.30 & 1.90 &  0.69$\pm$0.69 &  77$\pm$20.1 \\
 42 & --- & 5 41 49.29 & -1 52 44.3 & 17.81 & 15.87 & 14.34 &  1.02$\pm$1.45 & 0.00 & --- & 133$\pm$23.4 \\
 43 & --- & 5 41 37.40 & -1 52 44.6 & 13.41 & 11.87 & 11.10 &  3.27$\pm$0.10 & 1.50 &  2.18$\pm$0.07 & 167$\pm$0.9 \\
 44 & HPS4 & 5 41 36.74 & -1 52 44.9 & 13.93 & 12.69 & 12.21 &  2.94$\pm$0.18 & 0.62 &  4.77$\pm$0.29 &  11$\pm$1.7 \\
 45 & --- & 5 41 51.47 & -1 52 47.1 & 15.51 & 13.91 & 12.99 &  1.96$\pm$0.36 & 1.03 &  1.91$\pm$0.35 &  80$\pm$5.2 \\
 46 & --- & 5 41 48.23 & -1 52 48.3 & 16.50 & 14.53 & 13.50 &  2.44$\pm$0.45 & 2.16 &  1.13$\pm$0.21 &  53$\pm$5.1 \\
 47 & --- & 5 41 46.79 & -1 52 54.7 & 16.89 & 15.02 & 13.99 &  5.52$\pm$1.42 & 2.11 &  2.61$\pm$0.67 &  65$\pm$7.1 \\
 48 & --- & 5 41 49.83 & -1 52 56.1 & 15.63 & 13.11 & 11.67 &  1.56$\pm$0.12 & 2.24 &  0.69$\pm$0.05 &  65$\pm$2.2 \\
 49 & --- & 5 41 50.68 & -1 53 06.5 & 16.20 & 14.73 & 13.74 &  $<$0.86 & 0.72 &  $<$1.20 &  83$\pm$60.4 \\ 
 50 & --- & 5 41 33.86 & -1 53 08.7 & 13.71 & 12.47 & 11.81 &  2.50$\pm$0.17 & 1.20 &  2.08$\pm$0.14 & 136$\pm$1.9 \\
 51 & HPS5 & 5 41 38.25 & -1 53 09.1 & 12.00 & 10.07 &  9.17 &  8.45$\pm$0.04 & 1.90 &  4.45$\pm$0.02 &   4$\pm$0.2 \\
 52 & --- & 5 41 36.38 & -1 53 09.3 & 15.39 & 13.84 & 13.05 &  4.32$\pm$0.36 & 1.57 &  2.75$\pm$0.23 & 174$\pm$2.4 \\
 53 & HPS6 & 5 41 37.90 & -1 53 11.2 & 11.92 & 10.50 &  9.86 &  4.37$\pm$0.09 & 1.23 &  3.57$\pm$0.07 &   0$\pm$0.6 \\
 54 & --- & 5 41 50.88 & -1 53 12.8 & 15.63 & 13.99 & 12.70 &  1.75$\pm$1.12 & 0.00 & --- &  81$\pm$15.4 \\
 55 & --- & 5 41 37.98 & -1 53 18.7 & 15.93 & 14.09 & 12.99 &  2.51$\pm$1.14 & 1.30 &  1.93$\pm$0.88 &   7$\pm$11.8 \\
 56 & --- & 5 41 36.54 & -1 53 18.8 & 14.72 & 12.93 & 12.02 &  3.94$\pm$0.29 & 1.88 &  2.10$\pm$0.15 & 163$\pm$2.1 \\
 57 & HPS7 & 5 41 34.41 & -1 53 19.5 & 12.72 & 11.16 & 10.44 &  5.22$\pm$0.05 & 1.45 &  3.60$\pm$0.03 & 118$\pm$0.3 \\
 58 & --- & 5 41 30.25 & -1 53 20.3 & 14.96 & 13.64 & 12.91 &  2.99$\pm$0.21 & 1.35 &  2.22$\pm$0.16 & 141$\pm$2.0 \\
 59 & --- & 5 41 33.83 & -1 53 23.4 & 11.84 & 10.28 &  9.39 &  1.14$\pm$0.02 & 1.79 &  0.64$\pm$0.01 & 129$\pm$0.6 \\
 60 & --- & 5 41 36.30 & -1 53 24.4 & 16.39 & 14.42 & 13.21 &  2.27$\pm$0.72 & 1.44 &  1.57$\pm$0.50 & 166$\pm$8.6 \\
 61 & --- & 5 41 43.49 & -1 53 24.6 & 16.48 & 13.63 & 12.05 &  3.05$\pm$0.37 & 2.72 &  1.12$\pm$0.14 &  85$\pm$3.4 \\
 62 & --- & 5 41 30.89 & -1 53 24.6 & 14.21 & 12.97 & 12.29 &  3.68$\pm$0.11 & 1.24 &  2.97$\pm$0.09 & 131$\pm$0.9 \\
 63 & HPS8 & 5 41 38.14 & -1 53 25.5 & 17.27 & 15.27 & 14.20 & 10.10$\pm$3.24 & 2.25 &  4.49$\pm$1.44 & 172$\pm$8.7 \\
 64 & --- & 5 41 39.49 & -1 53 26.9 & 13.57 & 10.93 &  9.61 &  7.49$\pm$0.08 & 2.98 &  2.51$\pm$0.03 & 169$\pm$0.3 \\
 65 & --- & 5 41 49.67 & -1 53 27.0 & 14.67 & 12.82 & 11.87 &  2.50$\pm$0.39 & 1.97 &  1.27$\pm$0.20 & 101$\pm$4.4 \\
 66 & HPS9 & 5 41 38.32 & -1 53 28.4 & 16.56 & 14.29 & 13.18 & 11.06$\pm$1.15 & 2.47 &  4.47$\pm$0.47 & 179$\pm$3.0 \\
 67 & --- & 5 41 35.03 & -1 53 28.7 & 14.27 & 13.02 & 12.38 &  2.33$\pm$0.14 & 1.17 &  1.99$\pm$0.12 & 129$\pm$1.7 \\
 68 & --- & 5 41 43.92 & -1 53 29.2 & 17.82 & 14.17 & 12.32 &  8.32$\pm$0.87 & 4.35 &  1.92$\pm$0.20 &  63$\pm$3.0 \\
 69 & --- & 5 41 49.64 & -1 53 29.3 & 15.69 & 14.11 & 13.49 &  2.11$\pm$1.05 & 1.03 &  2.05$\pm$1.02 & 106$\pm$12.8 \\
 70 & --- & 5 41 49.38 & -1 53 31.1 & 14.88 & 13.03 & 12.04 &  3.20$\pm$0.37 & 2.04 &  1.57$\pm$0.18 &  75$\pm$3.3 \\
\hline
\end{tabular}
}
\end{center}
\begin{list}{}{}
\item[$^a$] {\scriptsize Source ID of highly polarized sources with $P_{H}/A_{H} > $ 3.38.} 
\item[$^b$] {\scriptsize Equatorial coordinates (J2000).} 
\item[$^c$] {\scriptsize $A_{H}$ estimated on the two color diagram.} 
\item[$^d$] {\scriptsize Errors are estimated using $\delta P_{H}$.} 
\end{list}{}{}

\setcounter{table}{1}
\clearpage
\begin{center}
\smallskip
Table 2: continued.
\smallskip
{\scriptsize
\begin{tabular}{ccccccccccc} \hline \hline
No. & HPS ID & R.A.  & Dec.        &  $J$ & $H$ & $K_{\rm s}$ & $P_{H}$ & $A_{H}$ & $P_{H}/A_{H}$  &  $\theta _{H}$ \\
      &              & (${}^{\rm h}\ {}^{\rm m}\ {}^{\rm s}$) & (${}^{\circ} \ ' \ ''$) & (mag) & (mag) & (mag) & ($\%$) & (mag) & 
	  ($\%$ mag${}^{-1}$)  &  (${}^{\circ}$)  \\
\hline
 71 & --- & 5 41 49.31 & -1 53 32.7 & 14.84 & 13.02 & 12.09 &  2.59$\pm$0.30 & 1.93 &  1.34$\pm$0.16 &  85$\pm$3.3 \\
 72 & --- & 5 41 38.33 & -1 53 33.0 & 13.44 & 12.89 & 12.53 &  0.94$\pm$0.27 & 0.00 & --- & 121$\pm$8.0 \\
 73 & --- & 5 41 39.50 & -1 53 33.6 & 15.18 & 12.54 & 11.10 &  6.95$\pm$0.15 & 3.22 &  2.15$\pm$0.05 & 169$\pm$0.6 \\
 74 & --- & 5 41 40.20 & -1 53 34.1 & 15.12 & 11.98 & 10.02 &  8.21$\pm$0.27 & 2.88 &  2.85$\pm$0.09 & 168$\pm$0.9 \\
 75 & --- & 5 41 40.06 & -1 53 35.5 & 15.40 & 12.59 & 11.16 & 10.06$\pm$0.36 & 3.22 &  3.13$\pm$0.11 & 169$\pm$1.0 \\
 76 & --- & 5 41 35.91 & -1 53 37.1 & 15.48 & 14.12 & 13.33 &  2.31$\pm$0.52 & 0.72 &  3.23$\pm$0.73 & 153$\pm$6.3 \\
 77 & --- & 5 41 34.02 & -1 53 38.8 & 15.28 & 14.01 & 13.31 &  2.15$\pm$0.36 & 1.28 &  1.68$\pm$0.28 & 150$\pm$4.7 \\
 78 & --- & 5 41 36.32 & -1 53 39.3 & 14.08 & 12.30 & 11.33 &  2.90$\pm$0.12 & 1.98 &  1.46$\pm$0.06 & 156$\pm$1.2 \\
 79 & --- & 5 41 36.94 & -1 53 39.3 & 14.78 & 12.78 & 11.60 &  2.44$\pm$0.23 & 1.52 &  1.61$\pm$0.15 & 164$\pm$2.7 \\
 80 & --- & 5 41 45.86 & -1 53 44.8 & 16.50 & 13.89 & 12.47 &  1.97$\pm$0.95 & 3.10 &  0.63$\pm$0.31 &  68$\pm$12.4 \\
 81 & --- & 5 41 51.17 & -1 53 45.9 & 15.35 & 13.51 & 12.50 &  5.67$\pm$0.44 & 2.07 &  2.74$\pm$0.21 &  92$\pm$2.2 \\
 82 & --- & 5 41 33.03 & -1 53 46.5 & 15.97 & 14.26 & 13.37 &  1.55$\pm$0.34 & 1.81 &  0.85$\pm$0.19 & 127$\pm$6.1 \\
 83 & --- & 5 41 46.21 & -1 53 46.7 & 14.06 & 11.25 & 10.00 &  2.33$\pm$0.10 & 2.68 &  0.87$\pm$0.04 &  82$\pm$1.2 \\
 84 & --- & 5 41 40.63 & -1 53 48.2 & 17.70 & 14.62 & 12.81 &  7.89$\pm$1.39 & 2.92 &  2.70$\pm$0.48 & 154$\pm$5.0 \\
 85 & HPS10 & 5 41 35.76 & -1 53 48.3 & 16.34 & 14.94 & 14.16 & 14.45$\pm$0.85 & 1.47 &  9.83$\pm$0.58 & 112$\pm$1.7 \\
 86 & --- & 5 41 48.61 & -1 53 49.5 & 15.09 & 12.64 & 11.47 &  4.60$\pm$0.22 & 2.58 &  1.78$\pm$0.08 &  64$\pm$1.4 \\
 87 & --- & 5 41 33.98 & -1 53 51.0 & 13.72 & 12.07 & 11.21 &  3.05$\pm$0.10 & 1.73 &  1.76$\pm$0.06 & 136$\pm$0.9 \\
 88 & --- & 5 41 43.91 & -1 53 51.3 & 16.85 & 14.80 & 13.15 &  0.89$\pm$0.83 & 0.00 & --- &  58$\pm$19.5 \\
 89 & --- & 5 41 56.80 & -1 53 52.1 & 14.06 & 11.98 & 11.06 &  4.67$\pm$0.03 & 1.73 &  2.70$\pm$0.02 &  51$\pm$0.2 \\
 90 & --- & 5 41 41.38 & -1 53 52.2 & 17.80 & 13.77 & 11.77 &  4.35$\pm$0.27 & 4.68 &  0.93$\pm$0.06 & 150$\pm$1.8 \\
 91 & HPS11 & 5 41 36.60 & -1 53 54.4 & 11.90 & 10.25 &  9.55 &  4.05$\pm$0.03 & 1.17 &  3.46$\pm$0.02 & 146$\pm$0.2 \\
 92 & HPS12 & 5 41 34.03 & -1 53 55.9 & 15.66 & 14.35 & 13.55 &  3.18$\pm$0.58 & 0.60 &  5.26$\pm$0.96 & 147$\pm$5.1 \\
 93 & HPS13 & 5 41 38.08 & -1 53 57.2 & 13.75 & 11.58 & 10.58 &  7.22$\pm$0.07 & 2.02 &  3.57$\pm$0.04 & 162$\pm$0.3 \\
 94 & HPS14 & 5 41 31.70 & -1 53 57.3 & 12.53 & 11.14 & 10.56 &  4.66$\pm$0.04 & 0.88 &  5.27$\pm$0.04 & 127$\pm$0.2 \\
 95 & HPS15 & 5 41 39.09 & -1 53 58.4 & 12.30 & 10.14 &  8.94 & 10.21$\pm$0.02 & 2.55 &  4.00$\pm$0.01 & 171$\pm$0.0 \\
 96 & --- & 5 41 36.82 & -1 53 58.9 & 12.69 & 11.03 & 10.16 &  3.47$\pm$0.04 & 1.74 &  1.99$\pm$0.02 & 140$\pm$0.3 \\
 97 & HPS16 & 5 41 39.25 & -1 54 02.2 & 16.50 & 14.41 & 13.14 & 10.81$\pm$0.51 & 1.61 &  6.70$\pm$0.32 & 178$\pm$1.3 \\
 98 & --- & 5 41 42.54 & -1 54 02.9 & 16.79 & 13.73 & 11.92 &  4.19$\pm$0.53 & 2.87 &  1.46$\pm$0.19 & 112$\pm$3.6 \\
 99 & HPS17 & 5 41 31.39 & -1 54 05.1 & 14.00 & 12.75 & 12.24 &  3.77$\pm$0.12 & 0.87 &  4.31$\pm$0.14 & 128$\pm$0.9 \\
100 & --- & 5 41 45.06 & -1 54 06.4 & 18.24 & 13.06 & 10.27 &  2.55$\pm$0.15 & 6.70 &  0.38$\pm$0.02 &  62$\pm$1.7 \\
101 & --- & 5 41 36.69 & -1 54 08.2 & 13.51 & 11.76 & 11.01 &  3.81$\pm$0.07 & 1.29 &  2.95$\pm$0.05 & 141$\pm$0.5 \\
102 & HPS18 & 5 41 49.08 & -1 54 08.6 & 14.25 & 12.98 & 12.16 &  5.13$\pm$0.27 & 0.51 & 10.12$\pm$0.53 & 106$\pm$1.5 \\
103 & --- & 5 41 42.49 & -1 54 09.8 & 15.06 & 12.63 & 11.41 &  2.07$\pm$0.46 & 2.69 &  0.77$\pm$0.17 & 131$\pm$6.2 \\
104 & --- & 5 41 45.90 & -1 54 11.1 & 17.13 & 13.14 & 11.07 &  6.55$\pm$0.19 & 4.85 &  1.35$\pm$0.04 &  76$\pm$0.8 \\
105 & HPS19 & 5 41 39.19 & -1 54 14.1 & 14.39 & 11.43 &  9.70 &  9.34$\pm$0.05 & 2.75 &  3.39$\pm$0.02 & 171$\pm$0.2 \\
106 & --- & 5 41 36.67 & -1 54 15.1 & 17.11 & 15.12 & 13.98 &  2.43$\pm$1.21 & 1.55 &  1.57$\pm$0.78 & 155$\pm$12.8 \\
107 & --- & 5 41 48.68 & -1 54 15.8 & 16.30 & 13.47 & 11.88 &  5.93$\pm$0.46 & 2.66 &  2.23$\pm$0.17 &  73$\pm$2.2 \\
108 & --- & 5 41 32.36 & -1 54 19.4 & 17.08 & 15.23 & 13.94 &  3.24$\pm$1.62 & 1.12 &  2.89$\pm$1.44 & 113$\pm$12.8 \\
109 & --- & 5 41 44.36 & -1 54 20.3 & 16.16 & 13.02 & 11.39 &  3.21$\pm$1.23 & 3.71 &  0.86$\pm$0.33 & 156$\pm$10.2 \\
110 & HPS20 & 5 41 35.83 & -1 54 22.1 & 16.37 & 14.71 & 13.61 &  4.45$\pm$1.82 & 0.94 &  4.76$\pm$1.95 & 150$\pm$10.8 \\
111 & --- & 5 41 45.23 & -1 54 22.9 & 16.64 & 13.14 & 11.29 &  $<$1.49 & 4.27 &  $<$0.35 &  17$\pm$31.5 \\ 
112 & --- & 5 41 41.98 & -1 54 24.0 & 15.37 & 12.27 & 10.47 &  5.71$\pm$0.20 & 2.96 &  1.93$\pm$0.07 & 132$\pm$1.0 \\
113 & --- & 5 41 36.24 & -1 54 24.2 & 12.18 & 10.20 &  8.90 &  4.19$\pm$0.02 & 1.34 &  3.13$\pm$0.02 & 121$\pm$0.1 \\
114 & --- & 5 41 37.63 & -1 54 24.6 & 13.00 & 11.26 & 10.49 &  3.01$\pm$0.09 & 1.48 &  2.03$\pm$0.06 & 148$\pm$0.8 \\
115 & --- & 5 41 38.48 & -1 54 25.6 & 12.01 & 11.31 & 11.15 &  0.74$\pm$0.29 & 0.00 & --- & 107$\pm$10.6 \\
116 & --- & 5 41 56.85 & -1 54 26.3 & 15.90 & 13.76 & 12.66 &  6.93$\pm$0.18 & 2.35 &  2.95$\pm$0.08 &  57$\pm$0.8 \\
117 & --- & 5 41 38.72 & -1 54 29.2 & 16.55 & 14.24 & 12.76 &  1.79$\pm$2.09 & 1.77 &  1.01$\pm$1.18 &   8$\pm$21.7 \\
118 & HPS21 & 5 41 31.40 & -1 54 34.7 & 16.02 & 14.28 & 13.21 &  7.43$\pm$0.75 & 1.15 &  6.46$\pm$0.65 & 130$\pm$2.9 \\
119 & --- & 5 41 44.84 & -1 54 35.8 & 15.97 & 12.79 & 10.91 &  3.44$\pm$0.65 & 3.06 &  1.13$\pm$0.21 &   8$\pm$5.3 \\
120 & --- & 5 41 42.81 & -1 54 36.2 & 14.74 & 11.95 & 10.63 &  2.40$\pm$0.20 & 2.88 &  0.83$\pm$0.07 & 130$\pm$2.4 \\
121 & --- & 5 41 41.49 & -1 54 39.3 & 15.91 & 12.71 & 11.00 &  5.28$\pm$0.47 & 3.89 &  1.36$\pm$0.12 & 128$\pm$2.5 \\
122 & HPS22 & 5 41 34.49 & -1 54 41.1 & 12.11 & 10.70 & 10.12 &  4.40$\pm$0.03 & 0.85 &  5.17$\pm$0.03 & 109$\pm$0.2 \\
123 & --- & 5 41 36.46 & -1 54 41.2 & 15.35 & 13.79 & 12.95 &  4.03$\pm$0.64 & 1.65 &  2.44$\pm$0.39 & 109$\pm$4.5 \\
124 & --- & 5 41 32.85 & -1 54 44.3 & 13.71 & 12.72 & 12.20 &  2.55$\pm$0.11 & 0.86 &  2.99$\pm$0.13 & 119$\pm$1.2 \\
125 & --- & 5 41 41.39 & -1 54 44.6 & 18.41 & 13.65 & 11.40 &  5.39$\pm$0.40 & 5.30 &  1.02$\pm$0.08 & 126$\pm$2.1 \\
126 & --- & 5 41 42.62 & -1 54 45.8 & 15.38 & 12.36 & 10.86 &  4.52$\pm$0.34 & 3.45 &  1.31$\pm$0.10 & 128$\pm$2.1 \\
127 & --- & 5 41 45.12 & -1 54 47.1 & 15.41 & 11.79 &  9.70 &  1.22$\pm$0.41 & 3.64 &  0.34$\pm$0.11 & 137$\pm$9.1 \\
128 & --- & 5 41 36.83 & -1 54 47.9 & 14.55 & 12.77 & 11.86 &  5.26$\pm$0.36 & 1.87 &  2.81$\pm$0.19 & 119$\pm$2.0 \\
129 & --- & 5 41 45.71 & -1 54 49.4 & 15.63 & 13.04 & 11.46 &  $<$1.01 & 2.22 &  $<$0.46 &  84$\pm$150.5 \\ 
130 & HPS23 & 5 41 34.57 & -1 54 50.0 & 16.69 & 15.34 & 14.57 &  4.97$\pm$3.27 & 0.70 &  7.10$\pm$4.67 & 111$\pm$15.7 \\
131 & --- & 5 41 51.11 & -1 54 52.3 & 15.85 & 14.06 & 12.94 &  2.38$\pm$2.07 & 1.17 &  2.03$\pm$1.77 &  70$\pm$18.8 \\
132 & HPS24 & 5 41 31.86 & -1 54 53.7 & 13.44 & 11.68 & 10.98 &  5.56$\pm$0.03 & 1.26 &  4.41$\pm$0.03 & 118$\pm$0.2 \\
133 & --- & 5 41 38.19 & -1 54 55.0 & 14.74 & 12.66 & 11.52 &  5.65$\pm$0.27 & 2.39 &  2.37$\pm$0.11 & 122$\pm$1.4 \\
134 & --- & 5 41 45.96 & -1 55 01.9 & 16.77 & 12.97 & 10.83 &  6.42$\pm$0.99 & 3.94 &  1.63$\pm$0.25 &  85$\pm$4.4 \\
135 & --- & 5 41 42.45 & -1 55 02.6 & 16.16 & 12.83 & 10.89 &  6.81$\pm$0.23 & 3.27 &  2.08$\pm$0.07 & 106$\pm$0.9 \\
136 & --- & 5 41 45.32 & -1 55 02.7 & 17.24 & 13.99 & 12.21 &  8.50$\pm$1.77 & 4.09 &  2.08$\pm$0.43 &  90$\pm$5.8 \\
137 & --- & 5 41 45.69 & -1 55 04.9 & 18.30 & 13.84 & 11.72 & 15.95$\pm$1.39 & 4.90 &  3.25$\pm$0.28 &  78$\pm$2.5 \\
138 & HPS25 & 5 41 50.97 & -1 55 07.0 & 12.47 & 10.46 &  9.26 &  8.09$\pm$0.11 & 1.52 &  5.32$\pm$0.07 &  57$\pm$0.4 \\
139 & --- & 5 41 55.78 & -1 55 09.3 & 12.86 & 11.77 & 11.28 &  2.35$\pm$0.09 & 0.88 &  2.69$\pm$0.10 & 101$\pm$1.0 \\
140 & --- & 5 41 42.20 & -1 55 10.0 & 12.82 & 11.28 & 10.53 &  4.70$\pm$0.06 & 1.50 &  3.14$\pm$0.04 &  98$\pm$0.4 \\
\hline
\end{tabular}
}
\end{center}

\setcounter{table}{1}
\clearpage
\begin{center}
\smallskip
Table 2: continued.
\smallskip
{\scriptsize
\begin{tabular}{ccccccccccc} \hline \hline
No. & HPS ID & R.A.  & Dec.        &  $J$ & $H$ & $K_{\rm s}$ & $P_{H}$ & $A_{H}$ & $P_{H}/A_{H}$  &  $\theta _{H}$ \\
      &              & (${}^{\rm h}\ {}^{\rm m}\ {}^{\rm s}$) & (${}^{\circ} \ ' \ ''$) & (mag) & (mag) & (mag) & ($\%$) & (mag) & 
	  ($\%$ mag${}^{-1}$)  &  (${}^{\circ}$)  \\
\hline
141 & HPS26 & 5 41 33.47 & -1 55 13.4 & 14.05 & 12.49 & 11.84 &  6.11$\pm$0.09 & 1.04 &  5.89$\pm$0.09 & 119$\pm$0.4 \\
142 & HPS27 & 5 41 53.88 & -1 55 16.3 & 14.61 & 11.88 & 10.33 & 13.45$\pm$0.27 & 2.51 &  5.36$\pm$0.11 &  80$\pm$0.6 \\
143 & --- & 5 41 52.09 & -1 55 18.2 & 13.07 & 11.99 & 11.48 &  1.77$\pm$0.61 & 0.88 &  2.01$\pm$0.69 &  72$\pm$9.3 \\
144 & HPS28 & 5 41 34.87 & -1 55 18.5 & 10.77 &  9.76 &  9.25 &  5.45$\pm$0.01 & 0.84 &  6.51$\pm$0.02 & 113$\pm$0.1 \\
145 & HPS29 & 5 41 31.94 & -1 55 18.6 & 12.45 & 10.74 &  9.90 &  7.39$\pm$0.04 & 1.73 &  4.26$\pm$0.03 & 121$\pm$0.2 \\
146 & --- & 5 41 40.60 & -1 55 18.6 & 15.91 & 12.83 & 11.26 &  5.30$\pm$0.35 & 3.57 &  1.49$\pm$0.10 & 137$\pm$1.9 \\
147 & HPS30 & 5 41 51.00 & -1 55 21.3 & 13.85 & 12.94 & 12.56 &  2.49$\pm$0.67 & 0.57 &  4.34$\pm$1.16 &  99$\pm$7.4 \\
148 & HPS31 & 5 41 44.40 & -1 55 22.8 & 13.69 & 11.43 & 10.02 &  7.00$\pm$0.10 & 1.78 &  3.94$\pm$0.05 & 104$\pm$0.4 \\
149 & --- & 5 41 44.33 & -1 55 24.9 & 14.52 & 12.37 & 11.30 &  5.33$\pm$0.16 & 2.32 &  2.30$\pm$0.07 & 119$\pm$0.9 \\
150 & HPS32 & 5 41 31.47 & -1 55 32.7 & 14.22 & 12.13 & 10.79 &  6.11$\pm$0.11 & 1.51 &  4.06$\pm$0.07 & 111$\pm$0.5 \\
151 & HPS33 & 5 41 38.22 & -1 55 36.3 & 12.84 & 11.54 & 11.03 &  5.38$\pm$0.10 & 0.69 &  7.84$\pm$0.15 & 117$\pm$0.5 \\
152 & HPS34 & 5 41 34.43 & -1 55 38.6 & 16.35 & 14.50 & 13.53 &  9.07$\pm$1.03 & 2.01 &  4.52$\pm$0.52 & 105$\pm$3.2 \\
153 & --- & 5 41 34.76 & -1 55 52.2 & 17.20 & 13.93 & 11.78 & 11.06$\pm$0.50 & 0.00 & --- &  94$\pm$1.3 \\
154 & --- & 5 41 56.68 & -1 55 52.6 & 13.52 & 12.41 & 11.95 &  2.25$\pm$0.12 & 0.71 &  3.15$\pm$0.17 &  73$\pm$1.5 \\
155 & HPS35 & 5 41 36.39 & -1 55 54.9 & 12.85 & 11.43 & 10.89 &  7.12$\pm$0.05 & 0.82 &  8.69$\pm$0.06 & 115$\pm$0.2 \\
156 & --- & 5 41 51.43 & -1 55 55.8 & 17.47 & 14.15 & 12.59 &  9.32$\pm$1.01 & 3.39 &  2.75$\pm$0.30 & 111$\pm$3.1 \\
157 & --- & 5 41 54.30 & -1 55 56.6 & 13.31 & 12.29 & 11.75 &  2.73$\pm$0.08 & 0.89 &  3.06$\pm$0.09 & 111$\pm$0.8 \\
158 & --- & 5 41 38.41 & -1 55 58.3 & 16.16 & 13.29 & 11.92 &  8.39$\pm$0.22 & 3.07 &  2.73$\pm$0.07 & 109$\pm$0.8 \\
159 & HPS36 & 5 41 38.77 & -1 56 02.5 & 14.09 & 12.10 & 11.01 & 10.69$\pm$0.10 & 2.26 &  4.73$\pm$0.04 & 112$\pm$0.3 \\
160 & HPS37 & 5 41 33.72 & -1 56 06.9 & 15.31 & 12.56 & 11.22 & 12.74$\pm$0.25 & 3.04 &  4.18$\pm$0.08 & 110$\pm$0.6 \\
161 & HPS38 & 5 41 35.43 & -1 56 15.5 & 13.72 & 13.05 & 12.81 &  0.78$\pm$0.21 & 0.23 &  3.40$\pm$0.90 & 103$\pm$7.3 \\
162 & HPS39 & 5 41 45.50 & -1 56 17.3 & 11.72 & 11.08 & 10.86 &  1.17$\pm$0.02 & 0.18 &  6.45$\pm$0.11 & 116$\pm$0.5 \\
163 & --- & 5 41 42.54 & -1 56 17.7 & 13.82 & 12.22 & 11.46 &  4.98$\pm$0.09 & 1.56 &  3.18$\pm$0.06 & 116$\pm$0.5 \\
164 & --- & 5 42 01.14 & -1 56 21.4 & 13.80 & 12.63 & 12.13 &  2.11$\pm$0.07 & 0.86 &  2.46$\pm$0.08 &  85$\pm$0.9 \\
165 & HPS40 & 5 41 35.81 & -1 56 22.3 & 16.06 & 13.08 & 11.37 & 13.79$\pm$0.42 & 2.85 &  4.84$\pm$0.15 & 118$\pm$0.9 \\
166 & --- & 5 41 39.06 & -1 56 26.5 & 16.42 & 13.14 & 11.42 &  6.79$\pm$0.54 & 3.94 &  1.72$\pm$0.14 &  77$\pm$2.3 \\
167 & HPS41 & 5 41 49.81 & -1 56 27.2 & 15.12 & 13.19 & 12.28 &  7.17$\pm$0.32 & 1.92 &  3.74$\pm$0.17 &  85$\pm$1.3 \\
168 & HPS42 & 5 41 52.93 & -1 56 29.6 & 18.00 & 14.78 & 12.87 & 25.17$\pm$0.57 & 3.07 &  8.20$\pm$0.19 &  89$\pm$0.7 \\
169 & --- & 5 41 54.98 & -1 56 31.2 & 16.62 & 13.53 & 11.89 &  8.82$\pm$0.16 & 3.71 &  2.37$\pm$0.04 &  91$\pm$0.5 \\
170 & HPS43 & 5 41 52.95 & -1 56 35.1 & 13.68 & 11.84 & 11.05 &  9.69$\pm$0.06 & 1.40 &  6.91$\pm$0.04 &  97$\pm$0.2 \\
171 & --- & 5 41 59.79 & -1 56 36.8 & 15.86 & 14.56 & 13.84 &  3.81$\pm$0.33 & 1.33 &  2.87$\pm$0.25 &  91$\pm$2.4 \\
172 & --- & 5 41 43.45 & -1 56 42.6 & 15.43 & 12.52 & 11.23 &  8.12$\pm$0.07 & 2.80 &  2.90$\pm$0.02 & 106$\pm$0.2 \\
173 & HPS44 & 5 41 53.41 & -1 56 48.1 & 15.71 & 14.22 & 13.27 &  4.20$\pm$0.28 & 0.79 &  5.31$\pm$0.35 &  96$\pm$1.9 \\
174 & HPS45 & 5 41 54.92 & -1 56 48.3 & 14.26 & 12.66 & 11.76 &  7.16$\pm$0.07 & 1.83 &  3.92$\pm$0.04 &  92$\pm$0.3 \\
175 & --- & 5 41 50.57 & -1 56 52.4 & 16.34 & 14.42 & 13.29 &  3.35$\pm$0.38 & 1.41 &  2.37$\pm$0.27 &  68$\pm$3.2 \\
176 & --- & 5 41 35.85 & -1 57 12.2 & 15.58 & 12.62 & 11.21 &  7.13$\pm$0.13 & 3.21 &  2.23$\pm$0.04 & 115$\pm$0.5 \\
177 & HPS46 & 5 41 52.33 & -1 57 12.2 & 15.74 & 14.33 & 13.52 &  8.20$\pm$0.30 & 1.61 &  5.10$\pm$0.18 &  94$\pm$1.0 \\
178 & --- & 5 41 45.05 & -1 57 13.2 & 17.36 & 16.73 & 16.25 &  7.49$\pm$1.81 & 0.00 & --- &   3$\pm$6.7 \\
179 & HPS47 & 5 41 52.16 & -1 57 17.4 & 12.97 & 11.19 & 10.27 &  7.56$\pm$0.03 & 1.90 &  3.98$\pm$0.01 & 107$\pm$0.1 \\
180 & HPS48 & 5 41 44.96 & -1 57 17.5 & 13.84 & 13.23 & 12.94 &  1.26$\pm$0.07 & 0.28 &  4.56$\pm$0.27 & 105$\pm$1.7 \\
181 & HPS49 & 5 41 54.72 & -1 57 27.3 & 12.88 & 11.74 & 11.29 &  5.21$\pm$0.03 & 0.71 &  7.35$\pm$0.04 &  91$\pm$0.1 \\
182 & HPS50 & 5 41 38.61 & -1 57 30.1 & 13.85 & 12.44 & 11.79 &  7.31$\pm$0.07 & 1.29 &  5.69$\pm$0.05 & 110$\pm$0.3 \\
183 & HPS51 & 5 41 50.04 & -1 57 30.6 & 13.43 & 11.96 & 11.28 &  4.50$\pm$0.04 & 1.33 &  3.39$\pm$0.03 &  74$\pm$0.2 \\
184 & --- & 5 41 45.62 & -1 57 33.1 & 14.19 & 13.45 & 12.92 &  2.14$\pm$0.10 & 0.00 & --- & 102$\pm$1.4 \\
185 & HPS52 & 5 41 44.20 & -1 57 35.3 & 15.43 & 13.47 & 12.43 &  9.71$\pm$0.15 & 2.18 &  4.45$\pm$0.07 &  90$\pm$0.4 \\
186 & HPS53 & 5 41 44.51 & -1 57 40.2 & 13.78 & 11.75 & 10.84 &  6.56$\pm$0.03 & 1.80 &  3.64$\pm$0.02 & 100$\pm$0.1 \\
187 & HPS54 & 5 41 36.56 & -1 57 41.3 & 15.41 & 13.19 & 12.18 &  8.28$\pm$0.19 & 1.99 &  4.16$\pm$0.10 & 108$\pm$0.7 \\
188 & HPS55 & 5 41 53.19 & -1 57 43.9 & 12.78 & 11.09 & 10.06 &  7.18$\pm$0.02 & 1.07 &  6.71$\pm$0.02 & 105$\pm$0.1 \\
189 & --- & 5 41 30.82 & -1 57 44.1 & 16.00 & 15.48 & 15.39 &  1.55$\pm$1.36 & 0.00 & --- & 170$\pm$18.9 \\
190 & HPS56 & 5 41 35.49 & -1 57 44.7 & 14.12 & 13.48 & 13.26 &  1.41$\pm$0.18 & 0.17 &  8.36$\pm$1.08 & 110$\pm$3.7 \\
191 & --- & 5 41 50.15 & -1 57 44.8 & 13.87 & 11.74 & 10.50 &  5.09$\pm$0.06 & 1.72 &  2.96$\pm$0.04 & 127$\pm$0.4 \\
192 & --- & 5 41 32.57 & -1 57 49.8 & 17.86 & 15.69 & 14.75 &  3.67$\pm$1.83 & 1.82 &  2.01$\pm$1.00 & 103$\pm$12.8 \\
193 & HPS57 & 5 41 53.71 & -1 57 50.2 & 15.87 & 14.35 & 13.45 &  6.07$\pm$0.36 & 0.89 &  6.84$\pm$0.40 & 108$\pm$1.7 \\
194 & --- & 5 41 41.65 & -1 57 54.6 & 12.47 & 10.32 &  9.05 &  1.50$\pm$0.02 & 1.70 &  0.88$\pm$0.01 &  85$\pm$0.3 \\
195 & --- & 5 41 31.48 & -1 57 57.0 & 16.45 & 14.77 & 13.97 &  3.58$\pm$0.74 & 1.65 &  2.17$\pm$0.45 & 109$\pm$5.8 \\
196 & --- & 5 41 32.75 & -1 57 57.3 & 12.29 & 11.54 & 11.36 &  1.46$\pm$0.04 & 0.00 & --- & 109$\pm$0.7 \\
197 & --- & 5 41 32.39 & -1 57 58.5 & 17.81 & 16.26 & 15.58 &  $<$2.31 & 1.22 &  $<$1.89 &  21$\pm$30.2 \\ 
198 & --- & 5 41 59.01 & -1 58 01.2 & 17.53 & 14.96 & 13.73 &  3.35$\pm$0.47 & 2.74 &  1.22$\pm$0.17 & 118$\pm$4.0 \\
199 & HPS58 & 5 41 50.62 & -1 58 04.1 & 14.29 & 13.37 & 12.79 &  3.45$\pm$0.09 & 0.12 & 29.80$\pm$0.76 &  88$\pm$0.7 \\
200 & --- & 5 41 50.06 & -1 58 08.1 & 15.88 & 15.01 & 14.53 &  2.34$\pm$0.40 & 0.73 &  3.21$\pm$0.54 & 125$\pm$4.8 \\
201 & --- & 5 41 57.75 & -1 58 09.9 & 16.93 & 15.19 & 14.40 &  3.54$\pm$0.55 & 1.60 &  2.21$\pm$0.35 & 108$\pm$4.4 \\
202 & HPS59 & 5 41 54.07 & -1 58 16.9 & 13.59 & 12.71 & 12.27 &  2.70$\pm$0.05 & 0.68 &  3.99$\pm$0.07 &  99$\pm$0.5 \\
203 & HPS60 & 5 41 38.17 & -1 58 17.6 & 14.16 & 12.50 & 11.86 &  6.32$\pm$0.06 & 1.13 &  5.61$\pm$0.06 & 106$\pm$0.3 \\
204 & HPS61 & 5 41 45.34 & -1 58 19.0 & 15.97 & 15.40 & 15.14 &  0.96$\pm$0.56 & 0.21 &  4.60$\pm$2.66 &  81$\pm$14.3 \\
205 & --- & 5 41 47.24 & -1 58 20.1 & 16.85 & 13.58 & 11.84 &  6.46$\pm$0.14 & 3.97 &  1.63$\pm$0.03 &  94$\pm$0.6 \\
206 & --- & 5 41 43.60 & -1 58 22.3 & 15.37 & 13.64 & 12.81 &  3.44$\pm$0.13 & 1.73 &  1.98$\pm$0.08 & 107$\pm$1.1 \\
207 & HPS62 & 5 41 33.14 & -1 58 23.9 & 17.18 & 15.87 & 15.13 &  9.57$\pm$1.83 & 1.38 &  6.96$\pm$1.33 & 120$\pm$5.4 \\
208 & --- & 5 41 52.07 & -1 58 32.4 & 16.42 & 15.51 & 14.61 & 12.58$\pm$0.74 & 0.00 & --- & 140$\pm$1.7 \\
209 & --- & 5 41 42.03 & -1 58 40.3 & 16.72 & 14.97 & 14.19 &  1.52$\pm$0.55 & 1.52 &  1.00$\pm$0.37 &  22$\pm$9.8 \\
210 & HPS63 & 5 41 43.99 & -1 58 45.4 & 14.94 & 13.41 & 12.43 &  3.03$\pm$0.14 & 0.82 &  3.68$\pm$0.17 &  91$\pm$1.3 \\
211 & HPS64 & 5 41 41.86 & -1 58 48.9 & 14.06 & 12.71 & 11.78 &  2.62$\pm$0.07 & 0.55 &  4.79$\pm$0.13 & 103$\pm$0.8 \\
\hline
\end{tabular}
}
\end{center}

\clearpage
\onecolumn

\begin{center}
\smallskip
Table. 3: Fractions of Highly Polarized Sources \\
\smallskip
{\small
\begin{tabular}{cccc} \hline \hline
             $H^a$    &  Fraction$^b$  & Number$^b$ & Total Number$^c$   \\
			  (mag)  &  ($\%$)     &     \\
\hline
9.5     &  $31 \pm 8.9$ \%    & 12  & 39 \\
10.5   &  $29 \pm 8.2$ \%    & 12  & 42 \\
11.5   &  $31 \pm 9.5$ \%    & 11  & 35 \\
12.5   &  $36 \pm 11.3$ \%    & 10  & 28 \\
13.5   &  $39 \pm 13.0$ \%    & 9    & 23 \\
\hline
\end{tabular}
}
\begin{list}{}{}
\item[$^a$] {\small Dereddened $H$ magnitudes. Each bin is separated by 1 mag.} 
\item[$^b$] {\small The fraction and the number of stars with $P_{H}/A_{H} > 3.38$ in each bin.} 
\item[$^c$] {\small Total number of stars in each bin.} 
\end{list}{}{}
\end{center}

\clearpage
\begin{center}
\smallskip
Table. 4: Highly Polarized Sources with Known Spectral Type
\smallskip
{\small
\begin{tabular}{lcccccc} \hline \hline
             Star Name$^a$    &  $P_{H}$$^b$    & $A_{H}$$^c$ & $P_{H}/A_{H}$       &   ${\theta}_{H}$$^d$    &   M subclass$^e$ &Mass$^f$  \\
			                &   ($\%$)     &     (mag)   &  ($\%$ mag${}^{-1}$)   &                       (${}^{\circ}$)  &    &   (${M}_{\odot}$) \\
\hline
FLMN\_J0541328-0151271 (57)$^g$ &  4.22$\pm$0.41 &  0.94 &  4.49$\pm$0.43 & 147$\pm$2.8 & 8.50 & 0.02 \\
FLMN\_J0541506-0158041 (25) &  3.45$\pm$0.09 &  0.61 &  5.65$\pm$0.14 &  88$\pm$0.7 & 7.75 & 0.03 \\
FLMN\_J0541456-0157332 (29) &  2.14$\pm$0.10 &  0.20 & 10.61$\pm$0.52 & 102$\pm$1.4 & 8.00 & 0.03 \\
FLMN\_J0541497-0156198 (55)$^g$ & 19.88$\pm$3.89 &  5.10 &  3.90$\pm$0.76 &  79$\pm$5.5 & 7.00 & 0.04 \\
FLMN\_J0541537-0157503 (52) &  6.07$\pm$0.36 &  1.58 &  3.85$\pm$0.23 & 108$\pm$1.7 & 6.50 & 0.06 \\
\hline
FLMN\_J0541523-0157123 (46) &  8.20$\pm$0.30 &  1.31 &  6.25$\pm$0.22 &  94$\pm$1.0 & 5.75 & 0.11 \\
FLMN\_J0541534-0156482 (51) &  4.20$\pm$0.28 &  1.02 &  4.12$\pm$0.27 &  96$\pm$1.9 & 4.00 & 0.23 \\
FLMN\_J0541314-0154347 (30) &  7.43$\pm$0.75 &  2.08 &  3.58$\pm$0.36 & 130$\pm$2.9 & 3.25 & 0.35 \\
FLMN\_J0541357-0153483 (61) & 14.45$\pm$0.85 &  1.17 & 12.39$\pm$0.73 & 112$\pm$1.7 & 2.75 & 0.37 \\
FLMN\_J0541344-0154409 (02) &  4.40$\pm$0.03 &  0.83 &  5.31$\pm$0.04 & 109$\pm$0.2 & 1.75 & 0.72 \\
\hline
\end{tabular}
}
\begin{list}{}{}
\item[$^{a, c, e-f}$] {\small Taken from Table 3 of Levine et al. (2004). The numbers in parentheses of column 1 denote the source serial number in their catalog. $A_{H}$ was calculated by multiplying originally cataloged $A_{V}$ by a factor 5.71.} 
\item[$^{b, d}$] {\small This work.} 
\item[$^{g}$] {\small Not cataloged in Table 2 but have available $P_{H}$ measurements.} 
\end{list}{}{}
\end{center}

\twocolumn

\clearpage
\onecolumn
\begin{landscape}
\begin{figure}[t]
    \centering
    \begin{minipage}[c]{1.00\columnwidth}
        \centering\includegraphics[width=\columnwidth]{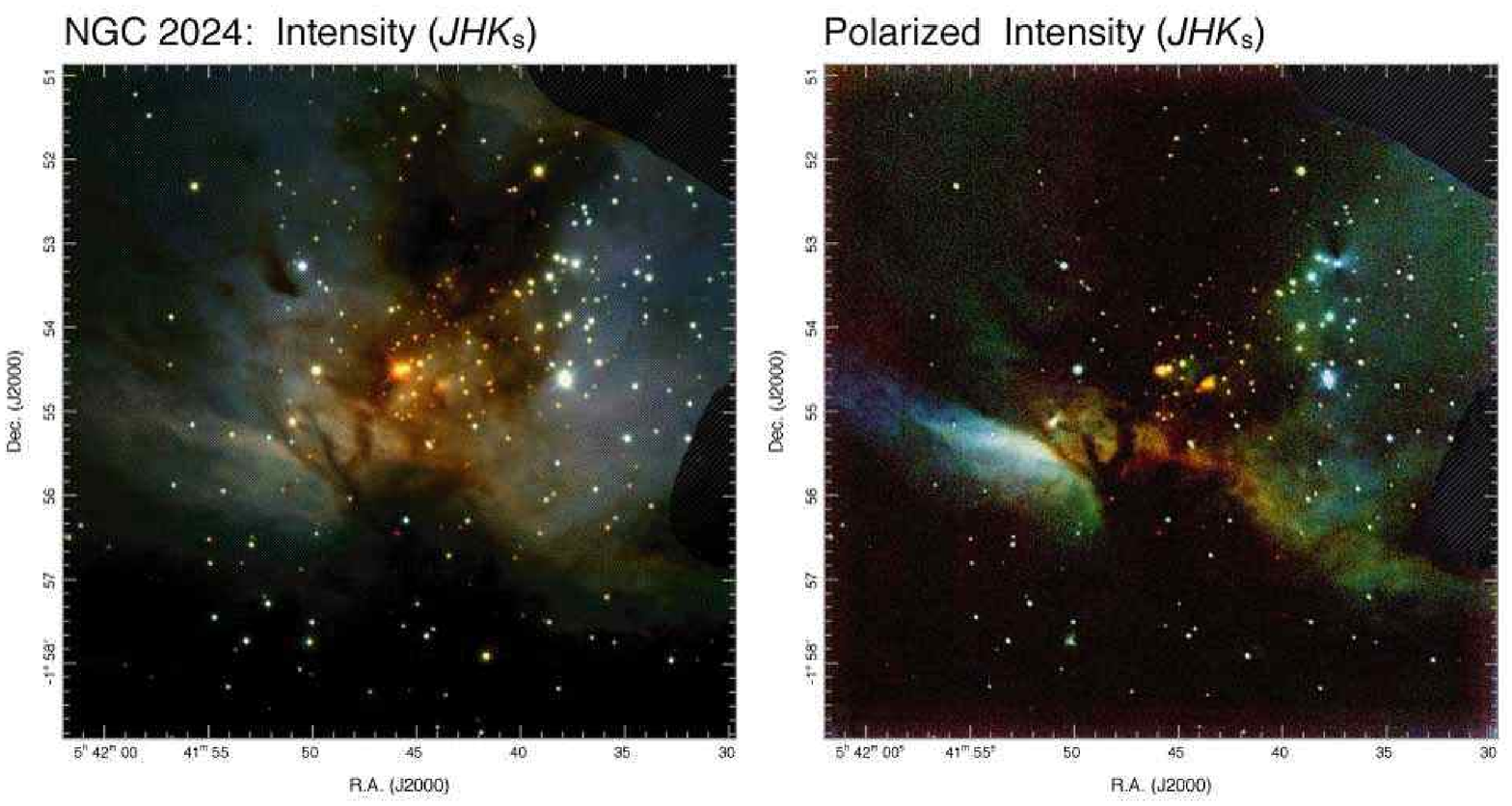}\\
    \end{minipage}
\smallskip
\\
\begin{flushleft}
Figure 1 $-$ Left: three-color composite of $JHK_{\rm s}$ intensity ($I$) images toward NGC 2024. Right: three-color composite of $JHK_{\rm s}$ polarized intensity ($PI$) images toward the same region. We note that there are bad pixel clusters on the $J$ image around the upper-right corner and the middle of right side, both of which are masked by hatched line. The images are in logarithmic scale.
\end{flushleft}
\end{figure}
\end{landscape}

\twocolumn

\clearpage
\onecolumn
\begin{figure}[t]
    \centering
    \begin{minipage}[c]{1.00\columnwidth}
        \centering\includegraphics[width=\columnwidth]{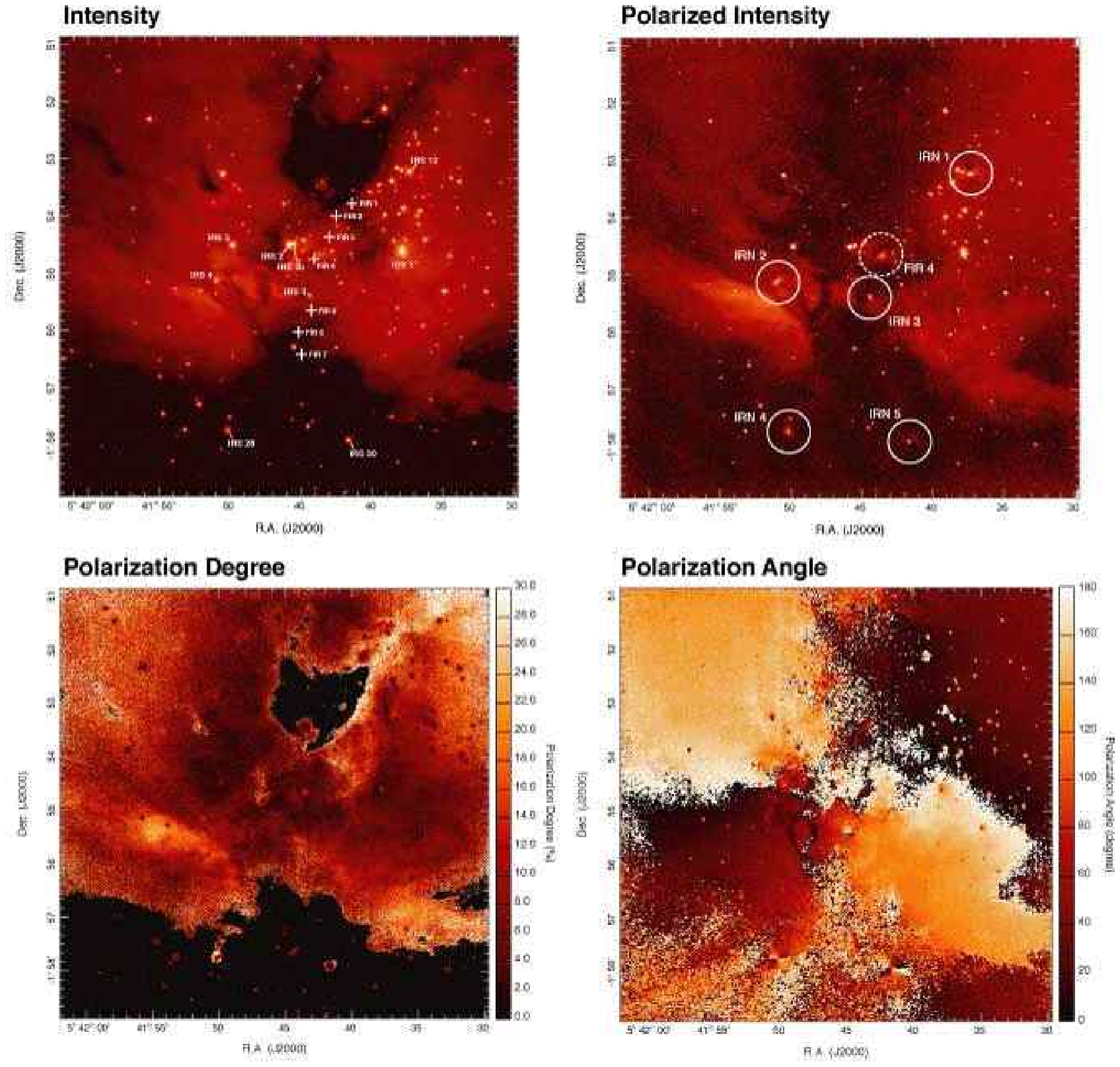}\\
    \end{minipage}
\smallskip
\\
\begin{flushleft}
Figure 2 $-$ Upper-left: Intensity ($I$) image of NGC 2024 in the $H$ band (logarithmic scale). Plus symbols denote far-infrared sources (FIR 1-6: Mezger et al. 1988; FIR 7: Mezger et al. 1992). Some of bright stars are marked with cataloged IRS numbers in Bernes (1989). Upper-right: Polarized intensity ($PI$) image of the same region (logarithmic scale). Newly found small infrared nebulae are enclosed by circles. A nebula enclosed by broken line circle was previously studied with infrared polarimetry (Moore \& Yamashita 1995). Lower-left: Polarization degree ($P$) image. $P_{H}$ is $\sim 30$ \% at maximum in the $H$ band. The regions with low S/N ($I/\delta I < 10$) are masked. Lower-right: Polarization angle ($\theta _{H}$) image. 
\end{flushleft}
\end{figure}

\clearpage
\begin{figure}[t]
    \centering
    \begin{minipage}[c]{1.00\columnwidth}
        \centering\includegraphics[width=\columnwidth]{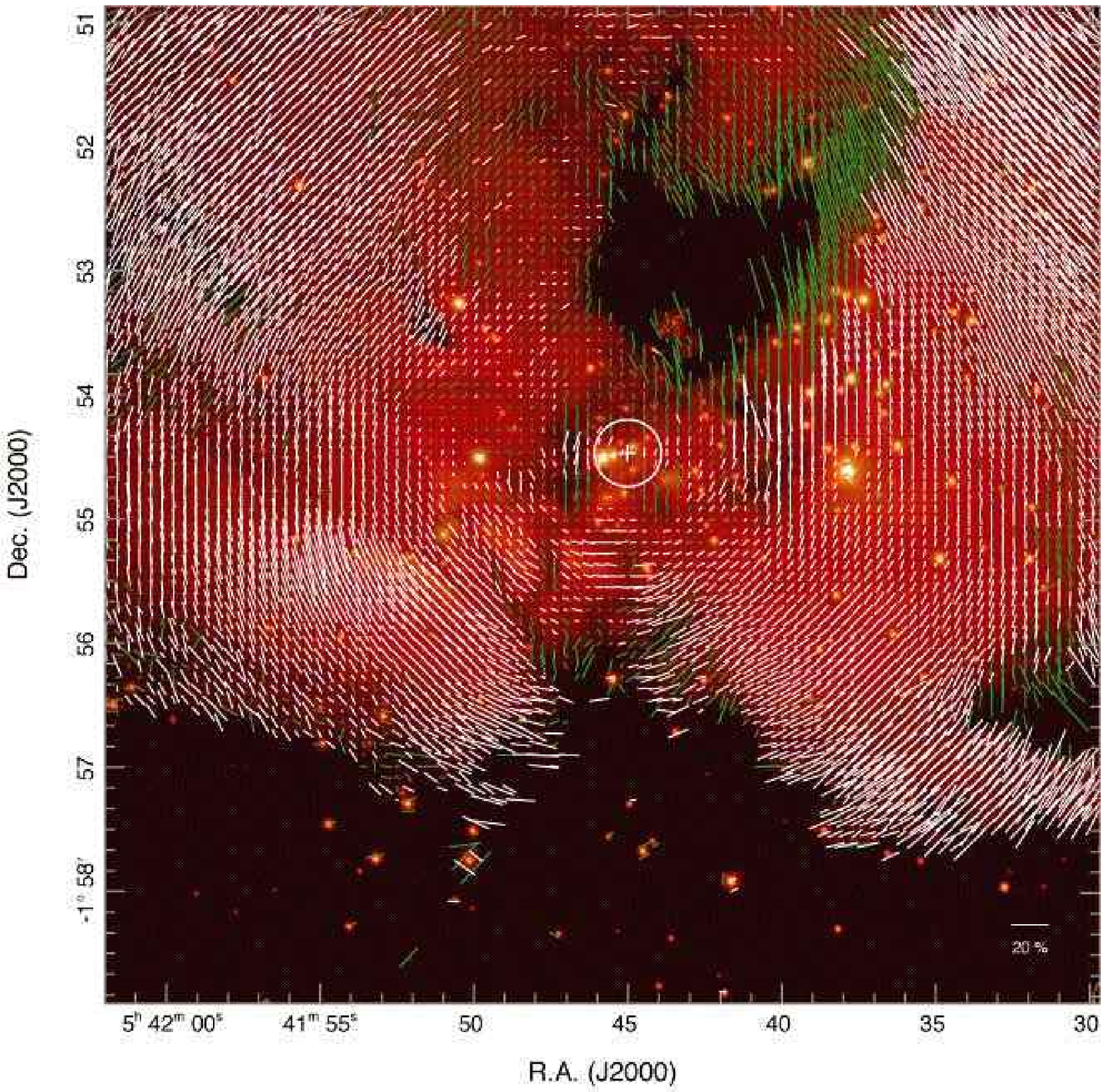}\\
    \end{minipage}
\smallskip
\\
\begin{flushleft}
Figure 3 $-$ Polarization vector map on the intensity image at $H$. White plus symbol indicate expected location of the illuminating source of NGC 2024. White circle around plus symbol denotes $1 \sigma$ error circle (see text). The polarization vectors on the low intensity region ($I/\delta I < 10$) are not plotted. 
\end{flushleft}
\end{figure}

\clearpage
\setcounter{figure}{3}
\begin{figure}[t]
    \centering
    \begin{minipage}[c]{0.85\columnwidth}
        \centering\includegraphics[width=\columnwidth]{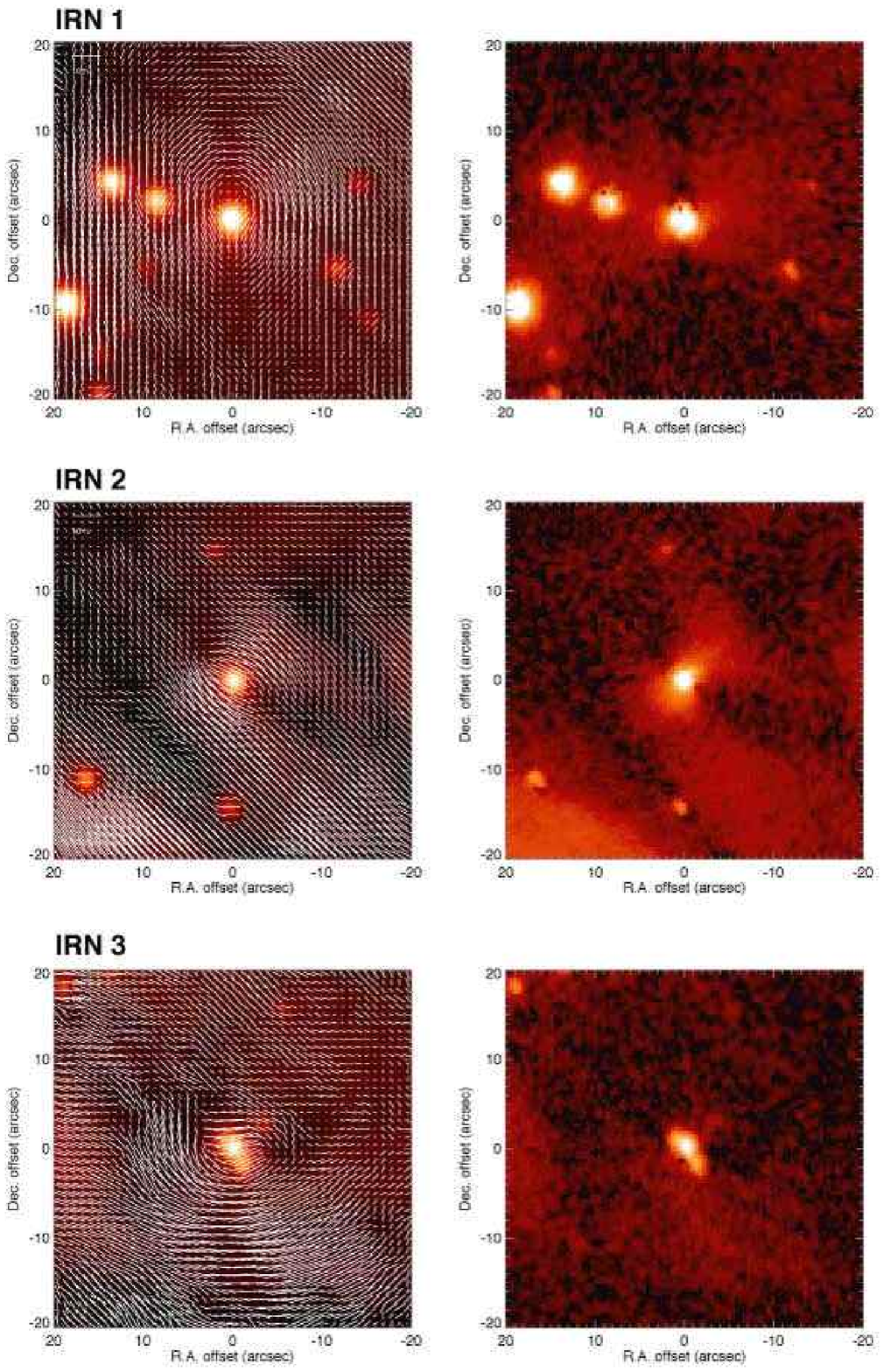}\\
    \end{minipage}
\smallskip
\\
\begin{flushleft}
Figure 4 $-$ Left column: Polarization vector maps on the $I$ images of small IRNe in NGC 2024. The polarization vectors on the low intensity region (IRN 1-3: $I/\delta I < 10$, IRN 4-5: $I/\delta I < 5$) are not plotted. Right column: Polarized intensity images. All the images are in the $H$ band and in logarithmic scale. 
\end{flushleft}
\end{figure}

\clearpage
\setcounter{figure}{3}
\begin{figure}[t]
    \centering
    \begin{minipage}[c]{0.85\columnwidth}
        \centering\includegraphics[width=\columnwidth]{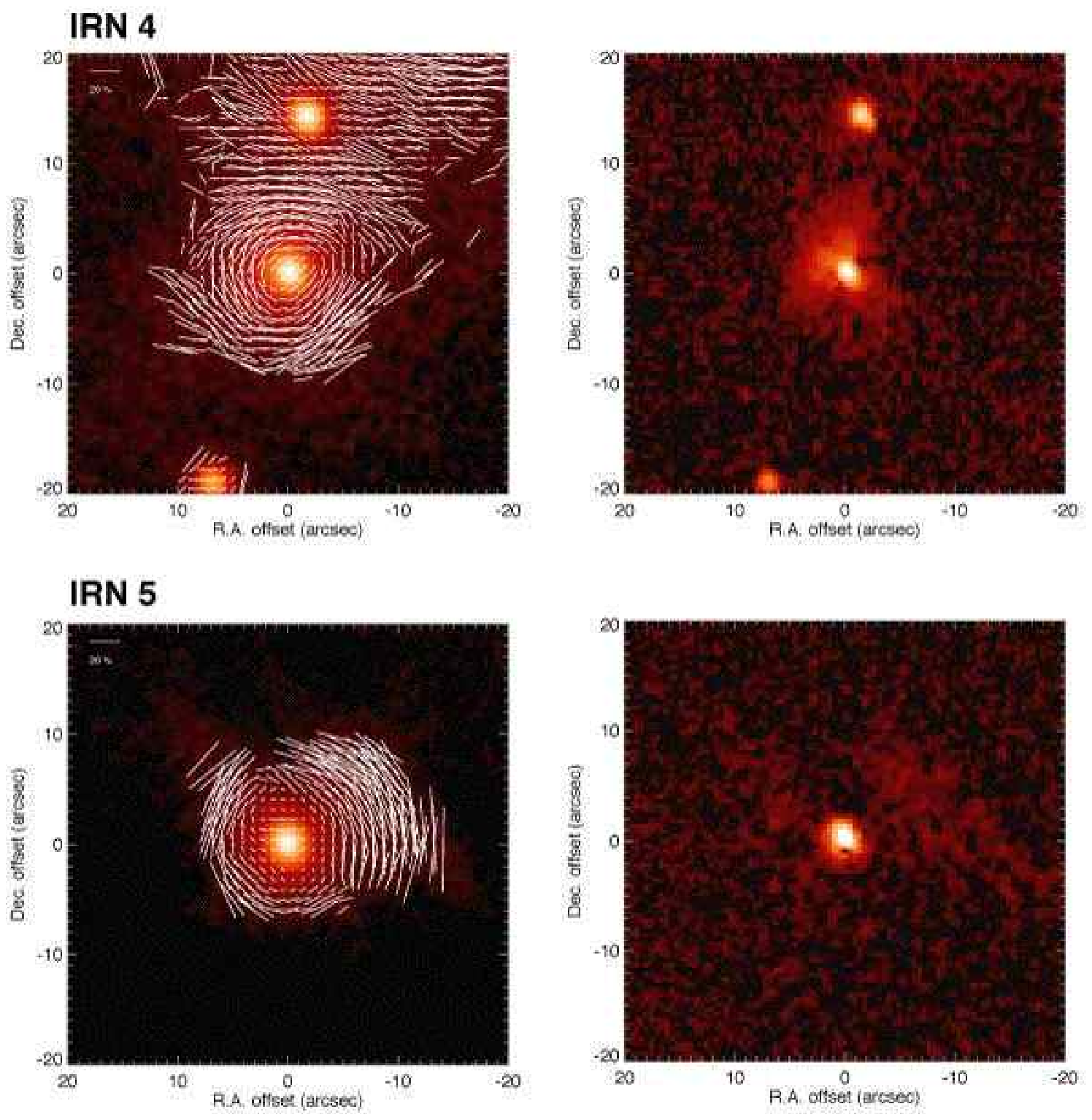}\\
    \end{minipage}
\smallskip
\\
\begin{flushleft}
Figure 4 $-$ continued.
\end{flushleft}
\end{figure}

\clearpage
\begin{figure}[t]
    \centering
    \begin{minipage}[c]{0.80\columnwidth}
        \centering\includegraphics[width=\columnwidth]{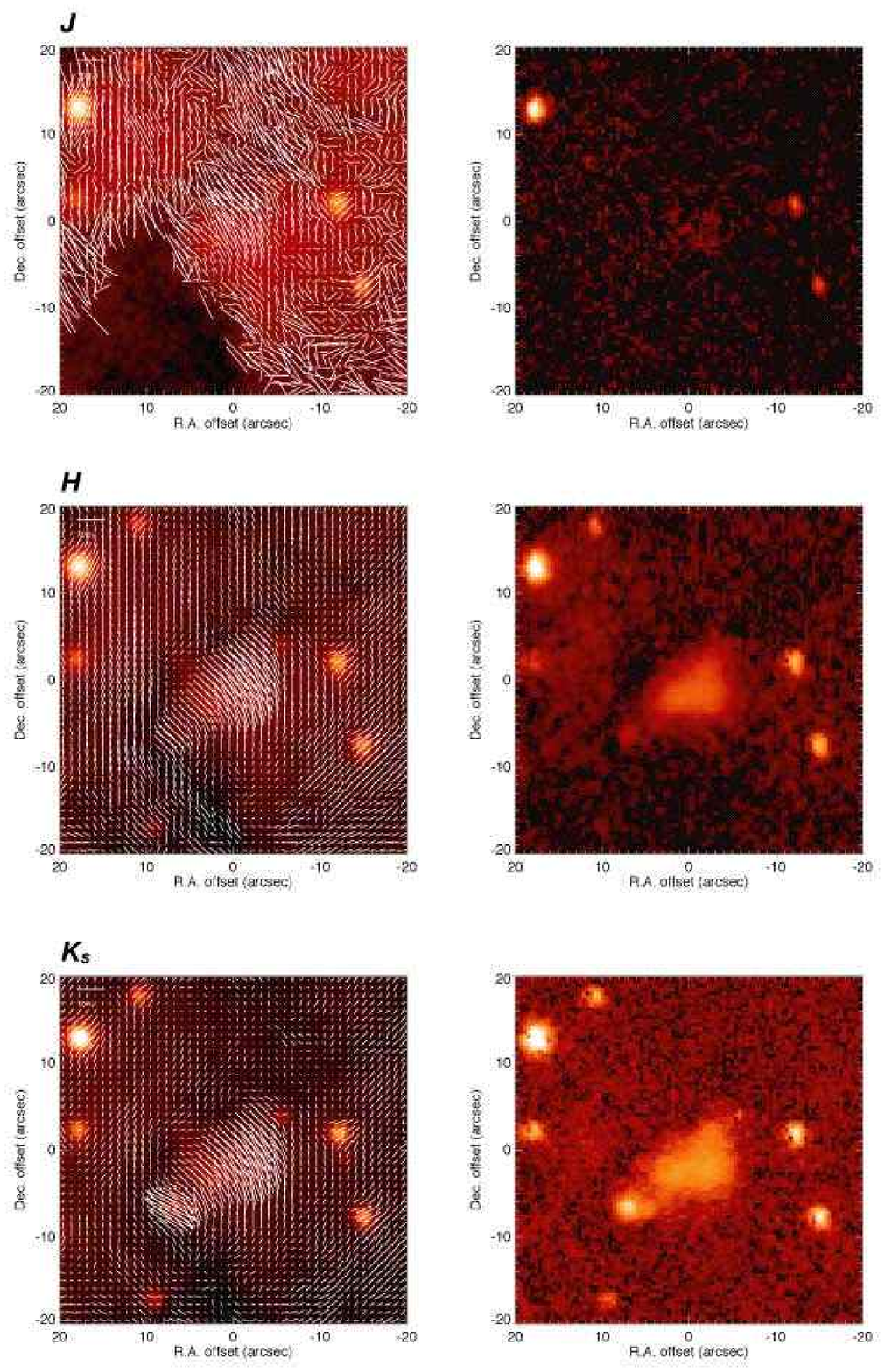}\\
    \end{minipage}
\smallskip
\\
\begin{flushleft}
Figure 5 $-$ Left column:  Polarization vector maps superposed on the $I$ images of $40'' \times 40''$ region around FIR 4. The polarization vectors on the low intensity region ($I/\delta I < 10$) are not plotted. Right column: Polarized intensity images. The images are in logarithmic scale. 
\end{flushleft}
\end{figure}

\clearpage
\begin{landscape}
\begin{figure}[t]
    \centering
    \begin{minipage}[c]{1.00\columnwidth}
        \centering\includegraphics[width=\columnwidth]{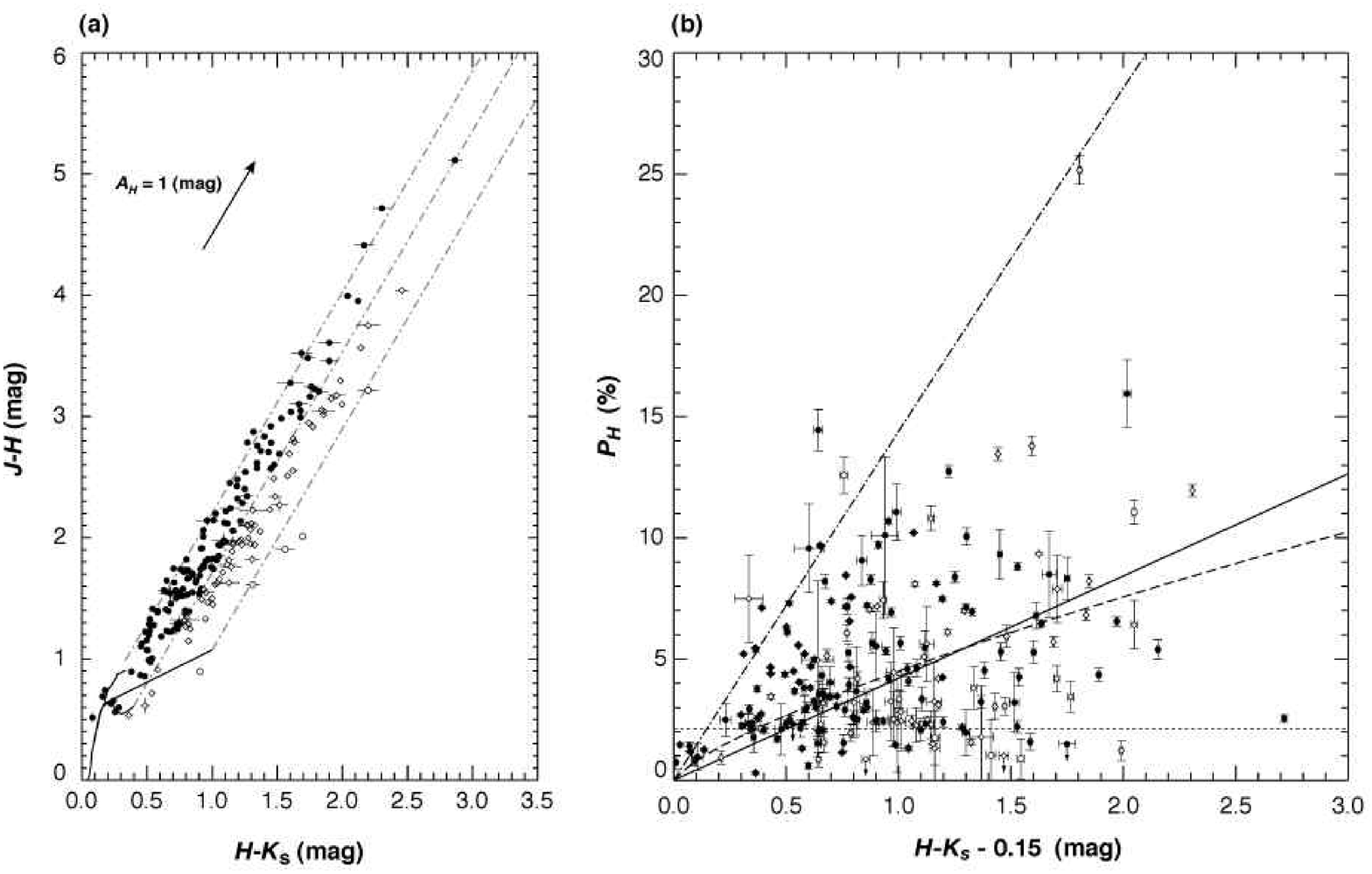}\\
    \end{minipage}
\smallskip
\\
\begin{flushleft}
Figure 6 $-$ Left panel: Two color diagram of the stars toward NGC 2024. Filled circles, open diamonds, and open circles denote dwarf+giant stars, PMS stars, and Class I candidates, respectively. The dwarf and giant locus were from Bessell \& Brett (1988), and T Tauri locus was from Meyer, Calvet, \& Hillenbrand (1997). The reddening vector was calculated using the reddening law from Nishiyama et al. (2006) and transformed from IRSF to 2MASS system. Right panel: $H-K_{\rm s} - 0.15$ vs. $P_{H}$ diagram. dashed and dot-dashed lines are the relationship for interstellar molecular clouds and its upper limit (Jones 1989). Solid line shows the result of linear fitting to all data points. Dotted  line shows the mode value of $P_{H}$. 
\end{flushleft}
\end{figure}
\end{landscape}

\clearpage
\begin{figure}[t]
    \centering
    \begin{minipage}[c]{0.70\columnwidth}
        \centering\includegraphics[width=\columnwidth]{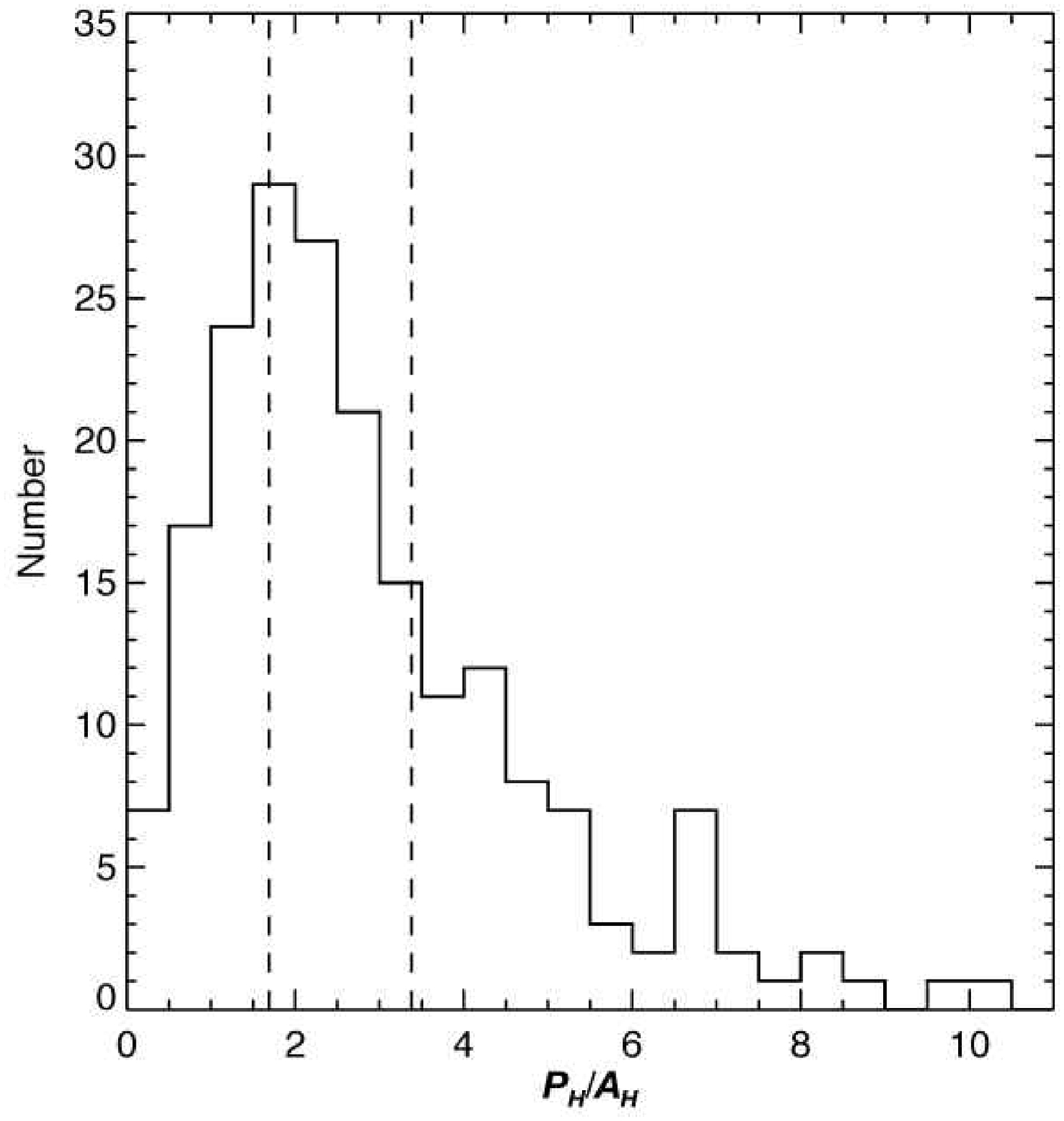}\\
    \end{minipage}
\smallskip
\\
\begin{flushleft}
Figure 7 $-$ Histogram of $P_{H}/A_{H}$ for all the stars except for protostar candidates. Dashed line on the left side shows $P_{H}/A_{H} = 1.69$, which corresponds to the solid line in Fig. 6b. Dashed line on the right side shows $P_{H}/A_{H} = 3.38$. 
\end{flushleft}
\end{figure}

\clearpage
\begin{figure}[t]
    \centering
    \begin{minipage}[c]{0.80\columnwidth}
        \centering\includegraphics[width=\columnwidth]{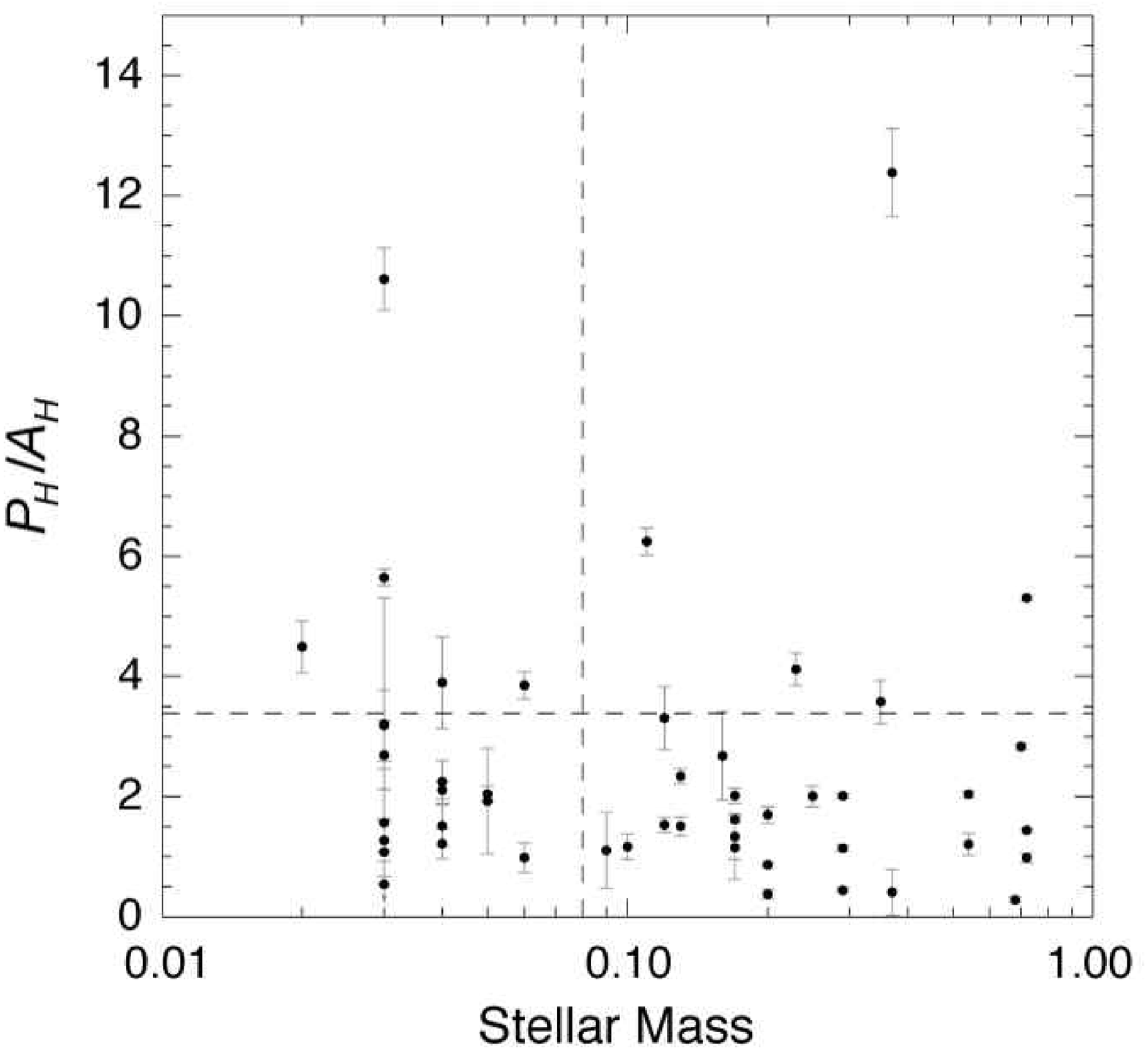}\\
    \end{minipage}
\smallskip
\\
\begin{flushleft}
Figure 8 $-$ Relationship between stellar mass and $P_{H}/A_{H}$ in NGC 2024. The value of stellar mass was taken from Levine et al. (2004). The mass range covers from brown dwarfs to solar-type stars. The vertical and horizontal broken lines denote the upper limit of brown dwarf mass (0.08 $M_{\odot}$) and $P_{H}/A_{H} = 3.38$, respectively. 
\end{flushleft}
\end{figure}

\clearpage
\begin{figure}[t]
    \centering
    \begin{minipage}[c]{1.00\columnwidth}
        \centering\includegraphics[width=\columnwidth]{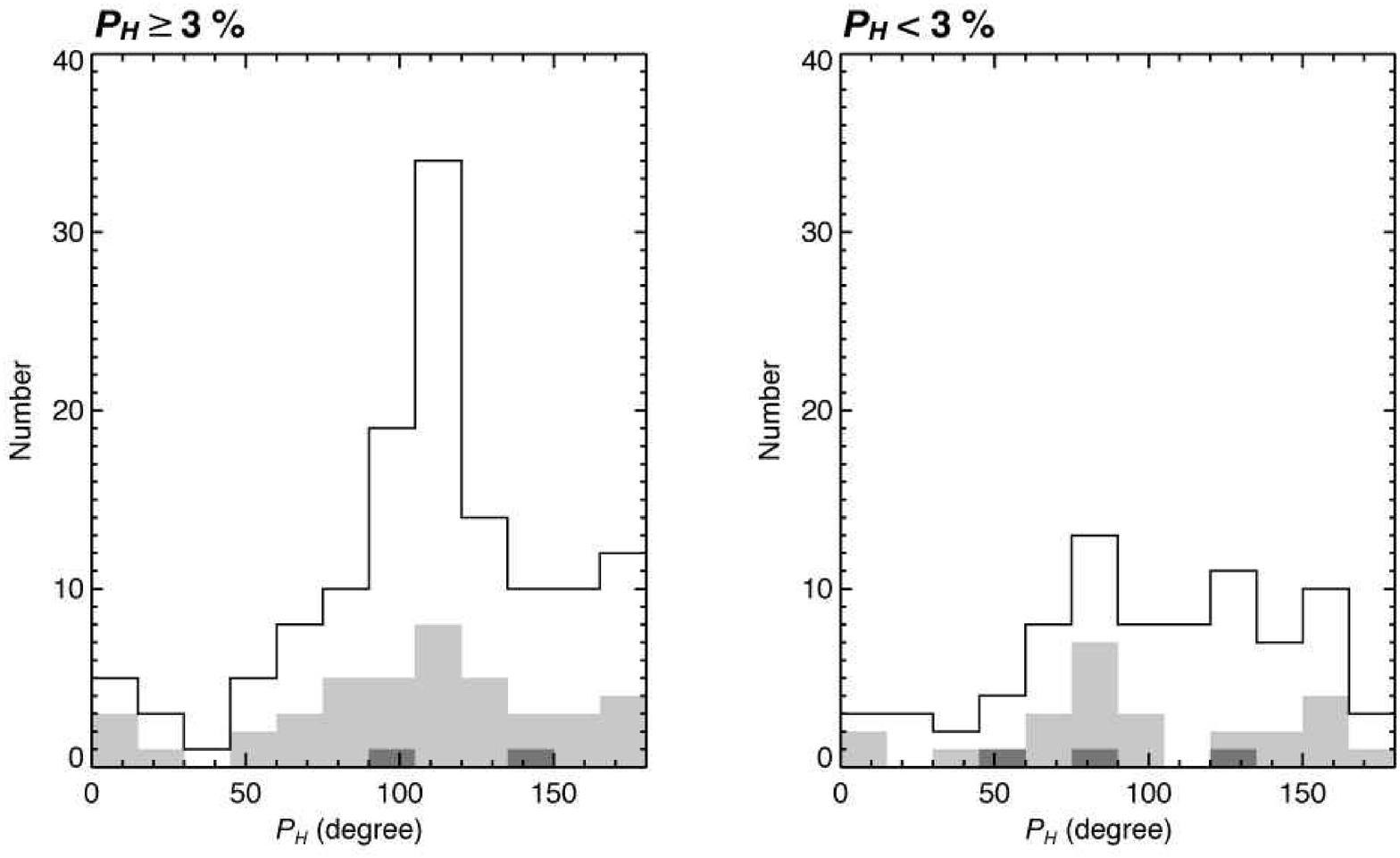}\\
    \end{minipage}
\smallskip
\\
\begin{flushleft}
Figure 9 $-$ Histograms of $P_{H}$ for the stars with $P_{H} \ge 3$ \% (left panel) and $P_{H} < 3$ \% (right panel). Dark gray, light gray and white colors show the number of protostar candidates, PMS stars, and dwarf+giant stars, respectively. 
\end{flushleft}
\end{figure}

\clearpage
\begin{figure}[t]
    \centering
    \begin{minipage}[c]{0.90\columnwidth}
        \centering\includegraphics[width=\columnwidth]{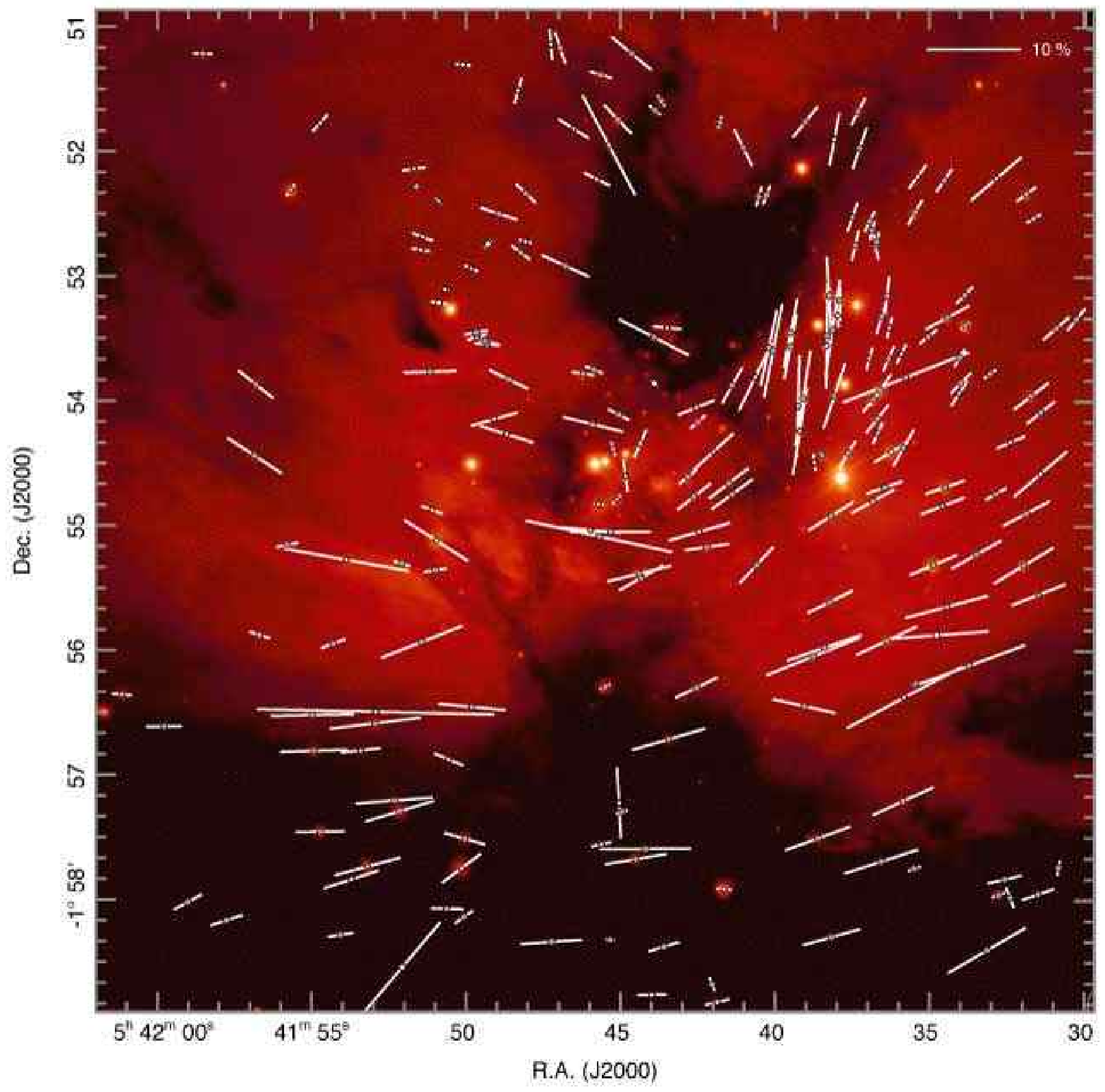}\\
    \end{minipage}
\smallskip
\\
\begin{flushleft}
Figure 10 $-$ Polarization vector of each stellar source on $H$ intensity image. Filled gray circles, open diamonds, and open circles denote dwarf+giant stars, PMS stars, and protostar candidates, respectively. 
\end{flushleft}
\end{figure}

\clearpage
\begin{figure}[t]
    \centering
    \begin{minipage}[c]{0.85\columnwidth}
        \centering\includegraphics[width=\columnwidth]{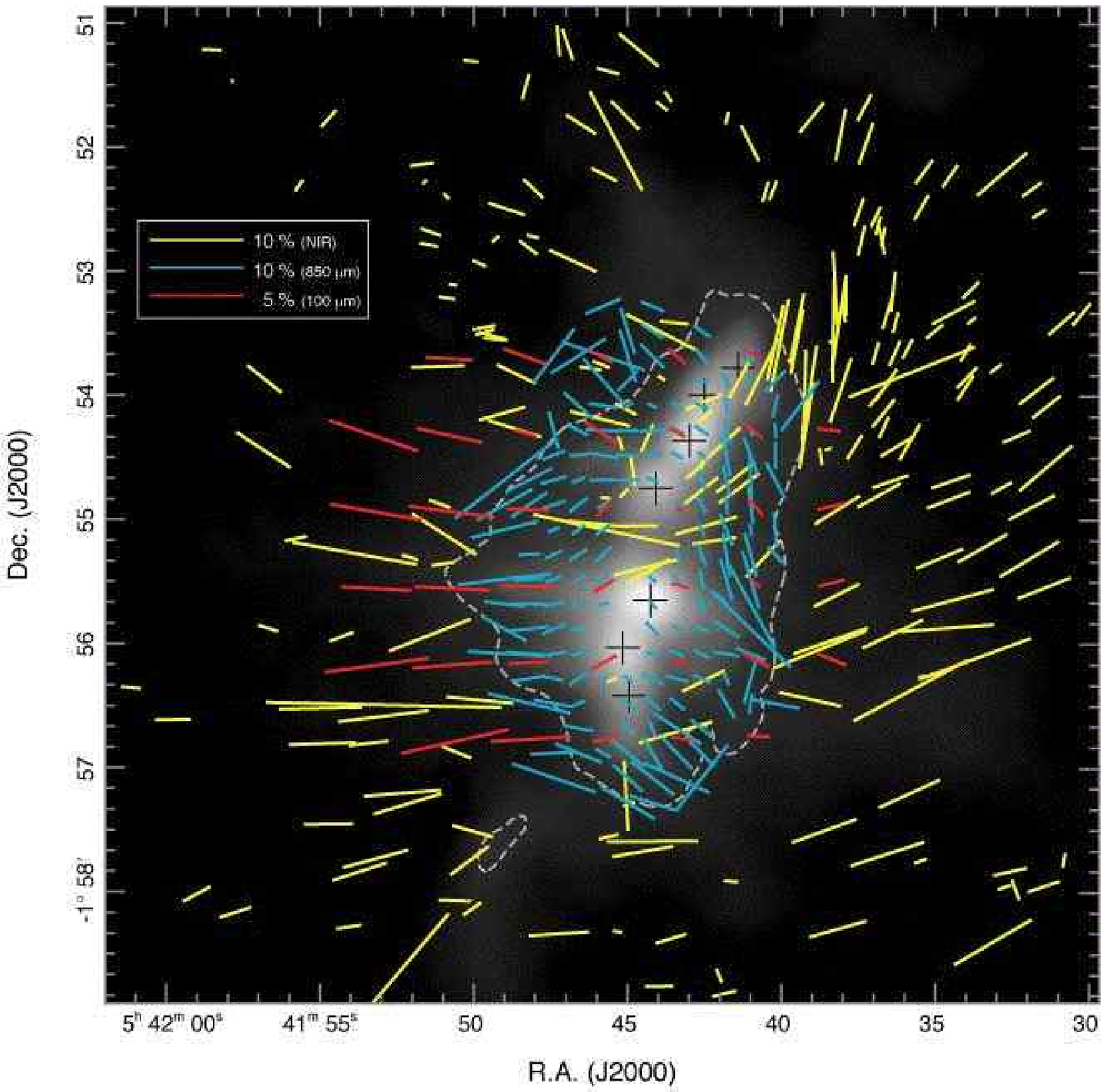}\\
    \end{minipage}
\smallskip
\\
\begin{flushleft}
Figure 11 $-$ Polarization vectors obtained through the near-infrared polarimetry at $H$ (yellow, this work), dust emission polarimetry at 850 $\mu$m (blue, Matthews, Fiege, \& Moriarty-Schieven 2002), and at 100 $\mu$m (red, Hildebrand et al. 1995; Dotson et al. 2000). Vectors from the dust emission polarimetry were rotated by 90 degrees for the comparison with the $E$-vector of dichroic polarization (inferred direction of magnetic field). The background image is the 850 $\mu$m dust continuum intensity map (logarithmic scale) kindly provided by Doug Johnstone. The contour of gray broken line denote the level of 0.9 Jy beam${}^{-1}$. Plus symbols denote far-infrared source, FIR 1-7 from north to south, reported by Mezger et al. (1988, 1992). 
\end{flushleft}
\end{figure}

\end{document}